\documentclass[12pt]{article} 

\usepackage{xr}
\usepackage[margin=1in]{geometry}
\usepackage[utf8]{inputenc}
\usepackage{graphicx}
\usepackage{authblk}
\usepackage{longtable}

\usepackage{xcite}
\usepackage{xr}

\makeatletter
\newcommand*{\addFileDependency}[1]{
  \typeout{(#1)}
  \@addtofilelist{#1}
  \IfFileExists{#1}{}{\typeout{No file #1.}}
}
\makeatother

\newcommand*{\myexternaldocument}[1]{%
    \externaldocument{#1}%
    \addFileDependency{#1.tex}%
    \addFileDependency{#1.aux}%
} 
\myexternaldocument{supp}

\usepackage{amsmath}
\usepackage{amssymb}
\usepackage{newtxmath}
\DeclareMathAlphabet{\mathpzc}{T1}{pzc}{m}{it}

\usepackage{bm}

\usepackage{caption}
\usepackage[dvipsnames]{xcolor}
\definecolor{dkgreen}{rgb}{0,0.6,0}
\definecolor{gray}{rgb}{0.5,0.5,0.5}
\definecolor{mauve}{rgb}{0.58,0,0.82}

\usepackage{xurl}

\usepackage{changepage}

\usepackage{setspace}
\usepackage[font=small,labelfont=bf]{caption}
\usepackage[utf8]{inputenc}

\title{Quantifying knowledge synchronisation in the 21st century} 
\author[1,2]{Jisung Yoon}
\author[3]{Jinseo Park}
\author[4]{Jinhyuk Yun \thanks{jinhyuk.yun@ssu.ac.kr}}
\author[1,5]{Woo-Sung Jung  \thanks{wsjung@postech.ac.kr}}

\affil[1]{Department of Industrial and Management Engineering, Pohang University of Science and Technology,
Pohang 37673, Republic of Korea.}
\affil[2]{Center for Complex Networks and Systems Research, Luddy School of Informatics, Computing, and Engineering, Indiana University, Bloomington, IN 47408, USA}
\affil[3]{Center for Global R\&D Data Analysis, Korea Institute of Science and Technology Information, Seoul 02456, Republic of Korea.}
\affil[4]{School of AI Convergence, Soongsil University, Seoul 06978, Republic of Korea.}
\affil[5]{Department of Physics, Pohang University of Science and Technology,
Pohang 37673, Republic of Korea.}

\date{\today}

\begin{document}

\maketitle

\begin{abstract}

Humans acquire and accumulate knowledge through language usage and eagerly exchange their knowledge for advancement. Although geographical barriers had previously limited communication, the emergence of information technology has opened new avenues for knowledge exchange. However, it is unclear which communication pathway is dominant in the 21st century.  Here, we explore the dominant path of knowledge diffusion in the 21st century using Wikipedia, the largest communal dataset. We evaluate the similarity of shared knowledge between population groups, distinguished based on their language usage. When population groups are more engaged with each other, their knowledge structure is more similar, where engagement is indicated by socio-economic connections, such as cultural, linguistic, and historical features. Moreover, geographical proximity is no longer a critical requirement for knowledge dissemination. Furthermore, we integrate our data into a mechanistic model to better understand the underlying mechanism and suggest that the knowledge ``Silk Road'' of the 21st century is based online.

\end{abstract}

\baselineskip24pt

\section*{Introduction}

Human language and knowledge are fundamentally intertwined and influence one another~\cite{code1980language}. Knowledge, which is defined as the ability to perceive and comprehend a subject, can be obtained through various sources, including memory, education, and practice \cite{grimm2014understanding}. Epistemologists traditionally investigated the nature and origins of knowledge. For instance, Immanuel Kant, a prominent epistemologist, claims that human perception is the basis of the general rules of nature that structure all our experiences~\cite{kant2000critique}. Because the experience could be different depending on the environments of population groups, knowledge structure can vary based on personalities, the country one lives in, or language profile based on a person’s social structure and education system. Humans conventionally acquire information through language, synthesizing knowledge from a flow of sensory experience~\cite{schieffelin1986language}. Thus, language profiles may influence the knowledge structure.

Researchers have considered that information can be spread through the mobility of people. For example, until the 16th century, the \textit{Silk Road} had played an important role in the transmission of knowledge between Europe and Asia~\cite{andrea2014silk, lu2016earliest}. Similarly, worldwide student exchanges have acted as a significant route for disseminating recent knowledge~\cite{bhandari2011global}. Active social interactions can aid efficient knowledge transmission between laborers at the team or company level~\cite{inkpen2005social, wu2007fostering, ringberg2008towards}. Language is also a crucial factor in the knowledge transfer process~\cite{welch2008importance, ambos2009impact}. Beyond physical contact between groups, modern information technology offers interactive online resources, such as user comments on a web page, social networks, and internet messengers, which allow knowledge to be transferred. Therefore, the emergence of information society raises intriguing questions: do social interactions sincerely influence the structure of human knowledge? If yes, what is the main contemporary channel of information distribution, which could be referred to as a contemporary \textit{Silk Road}? 

In this study, we attempt to answer the aforementioned questions using the knowledge structure for users of each language, who share their \textit{habitus} inherited from their antecedents. We compare the knowledge structures between languages. The way in which language is used reflects the innate knowledge structure of its users. Thus, we consider the usage of each language as a proxy for its users' \textit{habitus}, and hereafter, we use the term ``language'' to indicate the collective usage of the language users, unless otherwise specified. Researchers commonly use large scholarly databases as fundamental sources to explore knowledge structure for investigating the mechanisms of scientific innovations. However, these databases are suitable for examining shared knowledge within research communities rather than covering the society in general~\cite{qian2009knowledge, song2013detecting, su2010mapping, hu2014empirical,sakata2012identifying, fortunato2018science}. 

Although previous studies have achieved the quest of understanding human innovation to some degree, it also necessitates complementary data with more general coverage, including non-scholars. Wikipedia, on the other hand, enables us to construct knowledge structures encompassing general society. Wikipedia is an open online encyclopedia that is edited in real-time by contributors with hundreds of languages. Wikipedia is considered a representative example of collaborative knowledge, growing through collaboration and competition of contributors; and has been studied to understand the dynamics of collective intelligence~\cite{yasseri2012circadian, yasseri2012dynamics, yun2019early}. As a case study, the effect of cultural, linguistic, and regional factors on co-editing patterns is investigated along with the structure of global language networks~\cite{samoilenko2016linguistic, karimi2015mapping}. Researchers also show that one can extract geopolitical ties from the shared interest in Wikipedia's hyperlink structure~\cite{el2018capturing}. Several studies find that Wikipedia category data is a rich source of accurate knowledge~\cite{zesch2007analysis, nastase2008decoding, schonhofen2009identifying, ponzetto2009large, yoon_build_2018}, and thus, the Wikipedia category can be used as a good proxy for the knowledge structure by building flexible subject categories. In summary, Wikipedia is an abundant source of knowledge that one can use to examine the structure of knowledge in general society.

Here, we use Wikipedia’s multi-lingual linkage to evaluate the similarity of knowledge structures among different language editions to track the dominant pathway of contemporary information distribution. First, we construct 59 hierarchical knowledge networks based on the relationship between the pages and categories of each Wikipedia language edition, where each page or category is regarded as a scientific concept. Using a personalized page rank algorithm \cite{jeh2003scaling}, we build \emph{genealogy vectors} for each subject in the knowledge network. Then, using Wikipedia's multi-lingual linkage, we determine subject similarity by comparing the genealogical vectors of each subject among the knowledge networks from different language usage groups. We discover a plausible modular structure of languages comprising multidimensional factors, such as geographical, cultural, linguistic, and historical factors, by aggregating multiple topic similarities between languages into a knowledge structure similarity. Using this massive knowledge graph, we also discover geographically disassociated interactions, such as cooperative scientific research and social ties, by comparing with other socio-economic data, thereby explaining the synchronization of knowledge structures rather than geographical proximity. Furthermore, we successfully regenerate the similarity of empirical knowledge structures based on various socio-economic ties, supporting our previous observations, and uncover the potential mechanism underlying the synchronization of knowledge structures with the mechanistic model, inspired by the simple synchronization model~\cite{kuramoto1975international}. This study enables us to understand the contemporary \textit{Silk Road} of knowledge dissemination and that virtual social interactions shape the structure of human knowledge, as indicated by the massive records of online-based collaborative knowledge in the form of Wikipedia.

\section*{Result}

\subsection*{Knowledge structure of the languages and knowledge structure similarity.}

\begin{figure}[ht!]
    \centering
    \includegraphics[width=\textwidth]{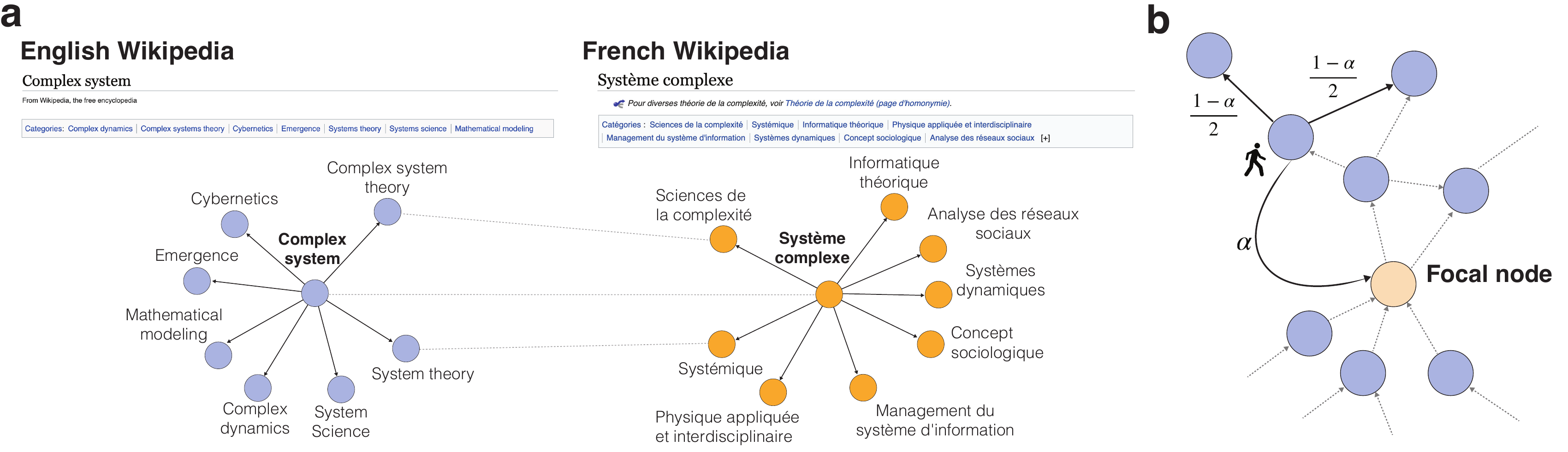}
    \caption{\textbf{Wikipedia knowledge network and genealogy vector of a subject.} 
    \textbf{a.} Example of a page-category hyperlink in English Wikipedia (left). The \texttt{page:Complex system} page has hyperlinks to several categories to which the page belongs. One can express such relations using the network (graph), where the node represents an entity (which can be a page, a category, or their union) with the links representing the hyperlink relationship between them. The identical \texttt{page:Complex system} page is titled \texttt{page:syst\`eme complexe} in French Wikipedia (right). Note that the hyperlink structures of two language editions are different, even for the identical entities. Between these two language editions, there are dotted and gray lines denoting the existence of language links between them, indicating that the two subjects are identical.
    \textbf{b.} Method for calculating the genealogy vector of a focal node. We obtain the genealogy vector using the personalized Page Rank algorithm, which calculates the probability that a random walker starting from the focal node visits other nodes. The random walker starts at the focal node and traverses with probability $1-\alpha$ to its nearest neighbors. The walker also occasionally returns to the focal node with a teleport probability, $\alpha$. The focal genealogy vector of the focal node is the random walker's stationary distribution for the visited nodes.
    }
    \label{fig:desc}
\end{figure}

From the linkage between Wikipedia pages and categories, we extracted a hierarchical \textit{knowledge network} of each language edition using the February 1, 2019, dump of fifty-nine Wikipedia language editions (see Methods). Categories are generally located at the end of a Wikipedia page and are designed to link related entries under a shared topic to make navigation easier. For the knowledge network, we designated a node as a category or page, treated as a proxy of a subject or scientific notion, and if there was a hyperlink from a subject to another subject, we assigned a directed link~\cite{yoon_build_2018}. We considered categories and pages to be identical when they shared an identical name (e.g., \texttt{category:science} and \texttt{page:science}); thus, we merged them into a single node with inheriting connected links. As an illustrative example displayed in Fig.~\ref{fig:desc}a, \texttt{page:Complex system} hyperlinks \texttt{category:System science}, and therefore, we assigned a directed link from the node \texttt{Complex system} to \texttt{System science} (for details, see Methods).

Our primary interest was to understand the dissemination of scientific and technological knowledge. We thus sampled a sub-network derived from an artificial root node named science \& technology, which is assigned as a common parent node of ``science'' and ``technology'' for each language edition. Note that ``Science'' covers all branches of science, such as applied sciences, formal sciences, natural sciences, and social sciences. A constructed knowledge network is directed and unweighted with several cycles from complex connections among nodes. Then, we obtained 59 knowledge networks based on their written language. The basic statistics of the derived networks are listed Table~\ref{supp:table:language_code}.

The obtained knowledge network represents the relationships between the subjects. As shown in Fig.~\ref{fig:desc}b, English Wikipedia users identify complex systems with complex dynamics and complex system theory, and other language users may consider different associations. For example, French Wikipedia users identify the complex system with distinct topics such as concept sociologique (sociological concepts) and analyse des r\'eseaux sociaux (social network analysis). To investigate such differences systematically, we introduced the concept of a \textit{genealogy vector} using personalized PageRank, which depicts how people correlate different subjects with the focal node regarding both nearest and non-nearest neighbors in the network (see Methods). Then, using multi-lingual linkage data, we compared genealogical vectors from different languages using \textit{subject similarity} (see Methods). We also considered the subject similarity as well as the cognitive similarities across different language editions of the same subject. Finally, we obtained a \textit{knowledge structure similarity} by aggregating the subject similarities between two language editions with averaging the similarity over all concepts.

\subsection*{Geographical proximity still influences, but socio-economic interaction shape the knowledge structure}

\begin{figure}[ht!]
    \centering
    \includegraphics[width=\textwidth]{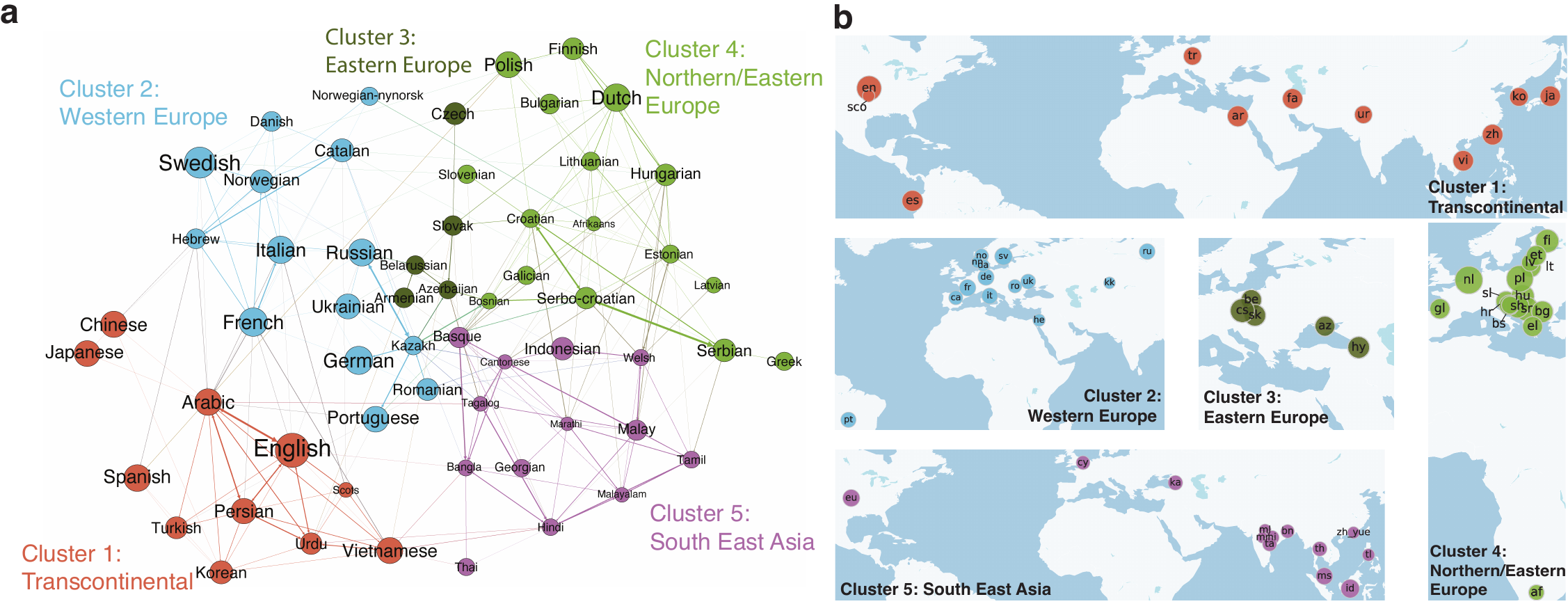}
    \caption{\textbf{Geographical proximity affects the similarity of knowledge structure across language usage groups.} \textbf{a.} We find five language communities from similarity networks using the Leiden algorithm \cite{traag2019louvain} (see Methods). The node colors indicate community memberships, whereas size indicates the number of documents of the corresponding Wikipedia edition on the log scale. \textbf{b.} Geographical dispersion of the languages for each community. The location of each language is estimated from the Wikipedia pageview data with geotags (see Supplementary Information for details).}
    \label{fig:network_map}
\end{figure}

A natural step forward was to find the possible sectors of languages whose members are more closely associated with each other. For this purpose, we constructed the \textit{similarity network} from the pairwise knowledge structure similarity, where nodes represent the language of Wikipedia, and the link’s weight indicates similarity between languages. Considering constructed similarity network is densely connected, we extracted the backbone of the networks by calculating the ego-centric importance of each link \cite{waltman2020principled} as follows 
\begin{equation}
    r_{ij}=\sum_{m,k} s_{mk} \frac{s_{ij}}{\sum_{k} s_{ik} \sum_{k} s_{kj}}.
\end{equation}
We then removed links whose normalized weight is under a certain threshold, $t$, and chose the minimum value of $t$ that network remains in a single component. As presented in Fig.~\ref{fig:network_map}a, five distinct communities are identified by the Leiden algorithm \cite{traag2019louvain} (see Methods) that indicate that the clusters seem to be affected by geographical proximity (Fig.~\ref{fig:network_map}b), which is similar to a previous study on Wikipedia bilateral ties. In this instance, geography best explains the formation of the cluster\cite{karimi2015mapping}. English is in the center and serves as a hub node, while intermediate hub languages such as Spanish, German, French, Russian, Portuguese, Chinese, and Dutch also function as cluster centroids~\cite{ronen2014links}. Four identified clusters generally correspond to languages spoken in Western Europe (light blue), Eastern Europe (dark green), Northern and Eastern Europe (light green), and Southeast Asia (purple). However, cultural, and historical backgrounds also play an important factor in the cluster, particularly for those of Northern and Eastern Europe. For example, Afrikaans, a language mostly spoken in South Africa, Namibia, and Botswana, evolved from European Dutch dialects \cite{pithouse2009making, heese1971herkoms} during the era of imperialism. One may note that geographic proximity does not appear to be a key determinant for Cluster 1: Transcontinental (orange), which spans the globe from the Far East to the Americas, and includes English as the de facto international language. These findings imply that knowledge distribution is still influenced by geographical proximity, which impacts the synchronization of knowledge structures between languages, while knowledge dissemination could also be influenced by other factors.

\begin{figure}[ht!]
    \centering
    \includegraphics[width=0.95\textwidth]{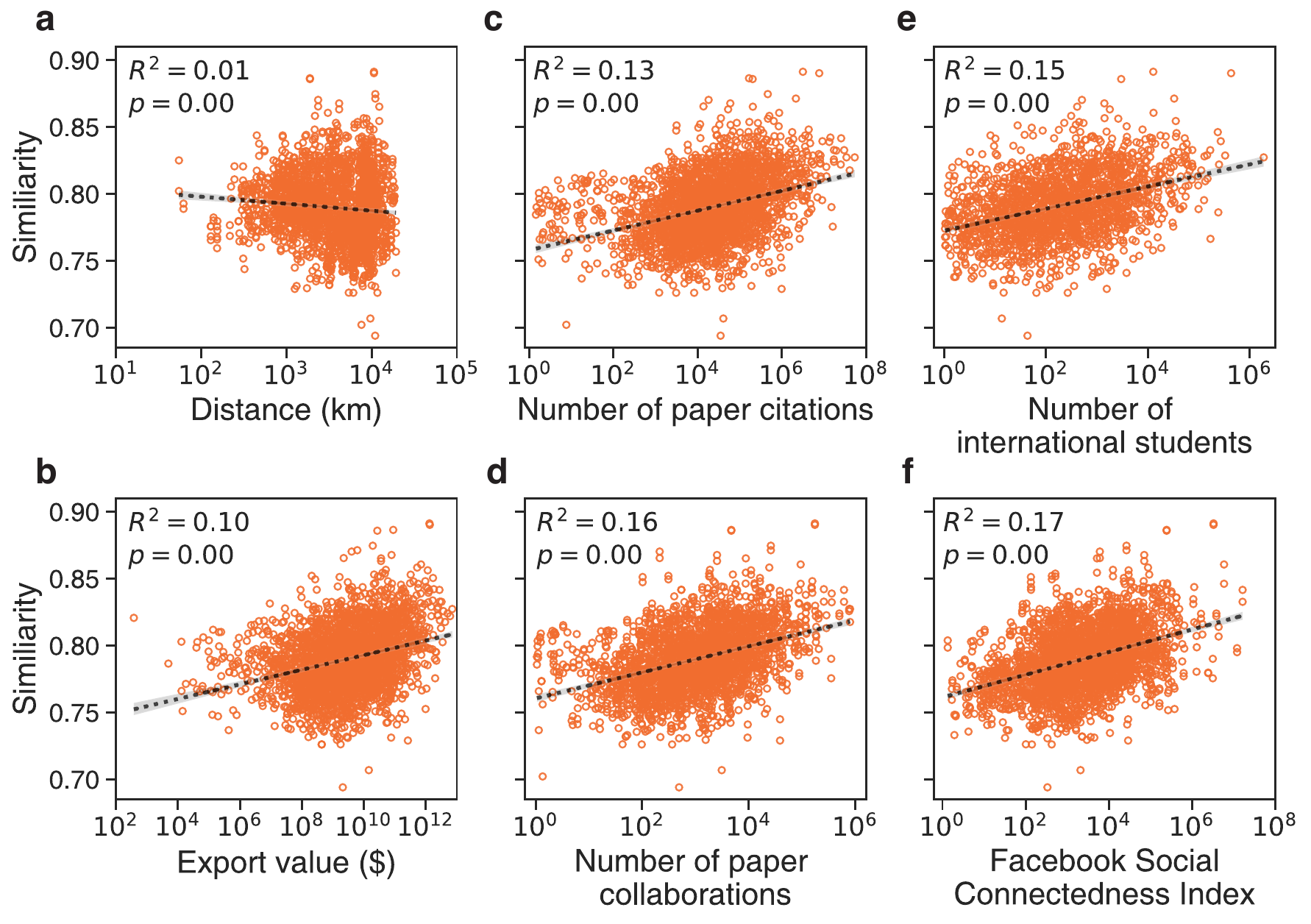}
    \caption{\textbf{Interrelationship of knowledge structure across language usage groups reveals the impact of socio-economic interactions.} The correlation between structural similarity of knowledge and socio-economic factors: \textbf{a.} geographical distance for the centroids of language pairs, \textbf{b.} Amount of exported goods for language pairs (IMF), \textbf{c.} Number of citations on papers for language pairs (SCOPUS), \textbf{d.} Number of co-authorship on paper for language pairs (SCOPUS), \textbf{e.} Number of the international students for language pairs (OECD), and \textbf{f.} Facebook Social Connected Index, the strength of connectedness between areas by represented by Facebook friendship ties, for language pairs. An increasing pattern of association is observed in the result.}
    \label{fig:regression}
\end{figure}

Nowadays, advances in technology provide new channels for interaction. For instance, modern information technology enables us to communicate with thousands, and even millions, of people in real time. One can also physically reach distant countries faster than ever before, with high-speed trains and air transportation being widely available. The cost of travel has also reduced significantly over time, owing to globalization \cite{hummels2007transportation}. Thus, such new routes can be new pathways for knowledge dissemination. Accordingly, we expand our analysis to include various language socio-economic connections to verify these new knowledge dissemination pathways. Because most socio-economic data focus on the interaction between countries, we first extract the language usages statistics of each country from the language database~\cite{ethnologue}. Then, we compare the projected socio-economic connection to a paired knowledge structure similarity, to identify potential contemporary Silk Roads for knowledge dissemination. First, we find that geographical distance no longer plays a central role in knowledge dissemination in the 21st century (Fig.~\ref{fig:regression}a). The geographical distance shows a weak and insignificant correlation with the knowledge similarities between countries (coefficient of determination is $R^2=0.01$). As we expect, there might be a new route, and the importance of the geographical proximity diminished in the knowledge exchange \cite{murray2020unsupervised}. 

We observe positive and more significant correlations from the non-geographical interactions (Fig.~\ref{fig:regression}). For example, the scientific interaction reflected in paper collaboration shows a higher coefficient of determination (Fig.~\ref{fig:regression}e; $R^2=0.16$) than those with geographical proximity (Fig.~\ref{fig:regression}a; $R^2=0.01$). Indirect scientific interactions also show a positive correlation, although comparably lower than direct interactions (Fig.~\ref{fig:regression}c; $R^2=0.13$). Because we consider the structure of knowledge denoted in Wikipedia under the science and technology topic, a strong connection in scientific collaboration will reasonably result in countries to have similar knowledge structures. Similarly, the soft power movement, which is counted as the number of international students, shows a high coefficient of determination because more than half of international students return to their homelands \cite{OECD2011} (Fig.~\ref{fig:regression}d; $R^2=0.15$). Furthermore, we find non-intellectual interactions positively correlate with knowledge similarity. For instance, the amount of export values, which are not directly related with the knowledge interchange, also correlates with knowledge structure to some degree (Fig.~\ref{fig:regression}b, $R^2=0.10$, IMF), and the result is robust for export values from different data sources (Fig.~\ref{suppfig:good_service} $R^2=0.09$, UN Comtrade). In other words, two language usage groups with strong socio-economic linkages are more likely to have similar knowledge structures. It is, nonetheless, a natural phenomenon because all these linkages are somehow related to knowledge exchange, which ultimately entails knowledge synchronization. 

By contrast, personal friendship is not necessarily associated with knowledge structure because it is not directly related to knowledge exchange; instead, it may be related to knowledge similarity through their information exchanges. Therefore, one might expect a weak or non-existent connection between friendship and knowledge. However, we find unanticipated significant and strong correlations between knowledge similarity and personal friendships, measured by the number of mutual friends in social media (Facebook social connectedness index (SCI); see Fig.~\ref{fig:regression}f, where their coefficient of determination $R^2=0.17$). This social link, which is reflected by the number of mutual friends in social networks, is the leading candidate for the Silk Road of the twenty-first century, which encompasses several levels of direct and indirect links among people on the web, although it is not widely considered the main channel of knowledge dissemination today.

In summary, we find the degree of association between socio-economic interactions and knowledge structure to occur in the following order: geographical distance (Fig.~\ref{fig:regression}a, $R^2=0.01$) $\ll$ export (Fig.~\ref{fig:regression}b, $R^2=0.10$) $<$ weak knowledge dissemination---paper citation (Fig.~\ref{fig:regression}c, $R^2=0.03$) $<$ soft-power movement---international students (Fig.~\ref{fig:regression}d, $R^2=0.15$) $<$ strong knowledge dissemination---paper collaboration (Fig.~\ref{fig:regression}e, $R^2=0.16$) $<$ mutual friendship on the web---Facebook SCI (Fig.~\ref{fig:regression}f, $R^2=0.17$). Taken together, the results demonstrate that social connections shape the collective knowledge structures of language users, regardless of whether it is explicitly connected to the knowledge transmission process. The possible mechanism behind the transmission will be discussed below, through our stochastic modeling.

\subsection*{Mechanistic model for the knowledge dissemination}

Our empirical analysis described in the previous sections reveals that i) knowledge structures are more likely to be similar if interactions exist between language usage groups and ii) the degree of association in knowledge structures varies based on the types of interactions. To understand the hidden mechanism of the observed correlation patterns, we identify the key factors driving the synchronization of knowledge structures. First, we assume that people are more likely to be similar when they interact more frequently and \textit{vice versa} \cite{gueguen2011similarity}. Second, the channel of interaction is progressively moving from a physical route to an online media space, which enables people to interact with overseas countries in real-time \cite{faraj2001web}. We consider a subject as a vector representation, which can be viewed as similar when they are close to each other. This is a similar concept to neural embedding \cite{peng2021neural}; however, we avoid declaring the embedding explicitly. Instead, we develop a mechanistic model to reproduce the synchronization of the knowledge structure with proximity among the language usage groups, using randomized initial vectors, motivated by the classic model to elucidate synchronization phenomena~\cite{kuramoto1975international}. 

By incorporating the aforementioned factors, we build a mechanistic model of knowledge spreading and synchronization. For simplicity, we only simulate the synchronization of a single subject's genealogy vector. The model comprises $N_{l}$ agents representing artificial language usage groups. Every agent has the capacity to store $d$ different subjects, and each digit represents their knowledge perception toward a target subject. Thus, each agent has a knowledge structure of $d$ dimensional vector similar to the genealogy vector discussed in the previous section. We then define the genealogy matrix, $V \in \mathcal{R}^{N_{l}\times d}$, by stacking the genealogy vectors of agents so that its row, $V_{i} \in \mathcal{R}^{d}$, represents the genealogy vector of agent $i$ (see Methods for the detail). We further assume that the initial status for agents is independent; thus, each row is orthogonal to the others. 
 
We additionally introduce proximity $p(i, j)$ from agent $i$ to agent $j$ to describe the degree of interaction between the agents. We use log-transformed empirical data (\textit{e.g.}, Facebook SCI and paper citations) as the proximity $p(i, j)$ and normalize them by dividing proximity $p(i, j)$ by the total sum of proximity for agent $i$, to use the proximity as the selection probability. Hence, the normalized proximity weight between agents $i$ to $j$ is given by $\hat{p}(i, j)=\frac{p(i, j)}{\sum_k{p(i,k)}}$.

For every simulation step, the genealogy vector of agent $i$ is updated as the following process. First, agent $i$ chooses its neighbor agent, $j$, with the probability of $\hat{p}(i, j)$, considering proximity. Then, the genealogy vector of agent $i$, $V_{i}$ is updated as follows:

  \begin{equation}
      	V_i(t + 1) =	V_i(t) + lr\cdot[V_{j}(t) - V_{i}(t)],
  \end{equation}
  
\noindent where $V_{j}(t)$ and $V_{i}(t)$ is the genealogy vector of group $i$ and chosen reference neighbor, $j$, at time $t$, respectively. $lr$ is the fixed learning rate for updating, and we use $lr=0.001$ for the results displayed in the main text. In this process, an agent’s pair with high proximity has a higher chance of being influenced and is more likely to become similar genealogy vectors as a consequence.

\begin{figure}[ht!]
    \centering
    \includegraphics[width=0.7\textwidth]{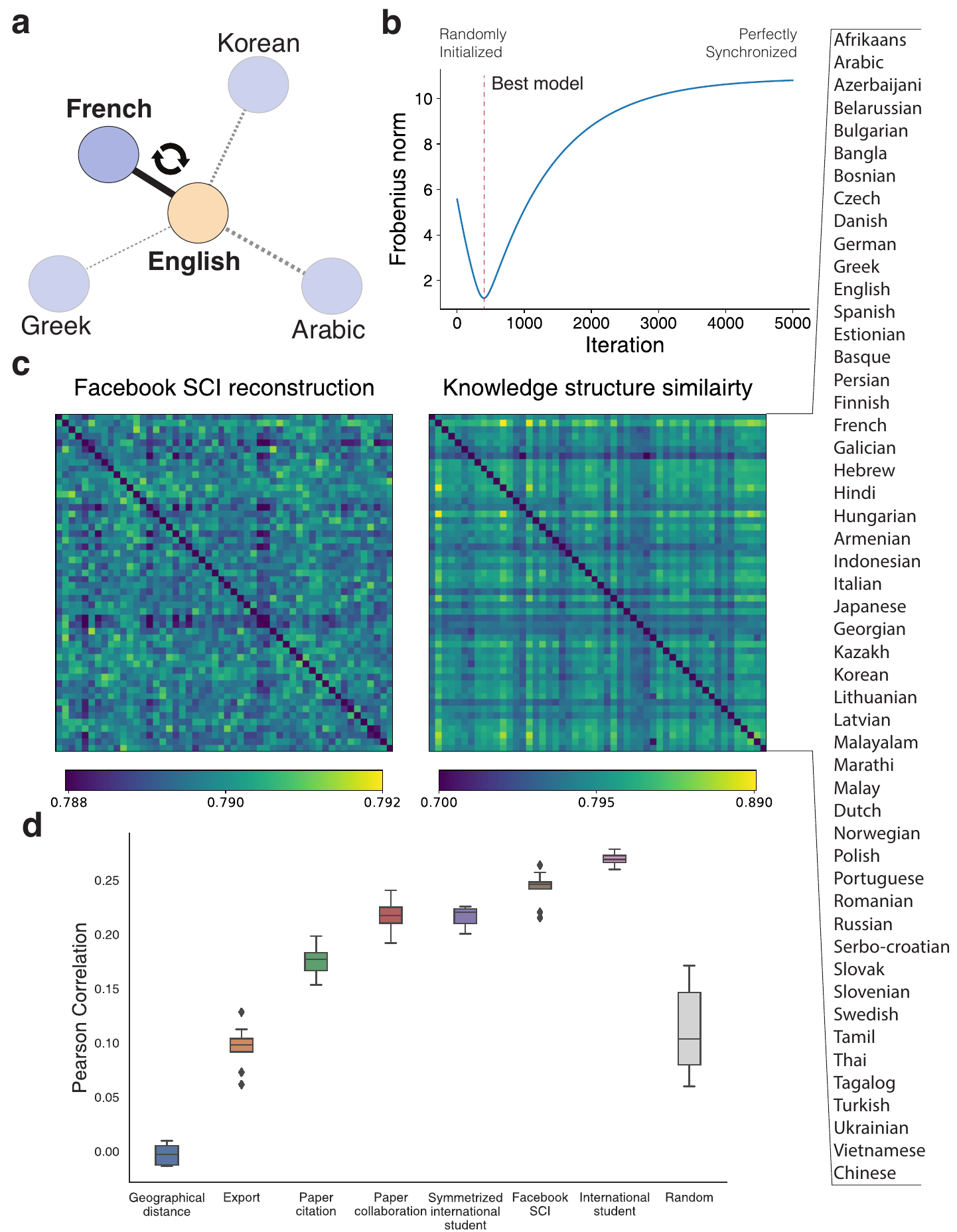}
    \caption{\textbf{Simple synchronization model for the knowledge similarity with various proximity indicators} \textbf{a.} Schematic diagram describing the model. For each time step, $t$, each agent selects one of their neighbors based on the probability proportional to the given proximity. As shown in the example, the English agent chooses the French agent as a neighbor for synchronization. \textbf{b.} Criteria for choosing the best model. The model begins with the randomly initialized orthogonal vectors, which are synchronized as more iterations occur. For each proximity, we choose the best model similarity at time step $t$ that shows the minimum Frobenius distance from the empirical knowledge similarity matrix. \textbf{c.} As an illustrative example, we present the snapshot of the best model similarity matrix for Facebook Social Connected Index ($r=0.257$, left) and empirical knowledge structure similarity from Wikipedia (right). \textbf{d.} The Pearson correlation between the best model similarity and empirical knowledge similarity indicates how well each proximity can reproduce the empirical knowledge similarity. To compensate for any randomness impact, we test 100 different initializations, repeated 10 times for each.
    }
    \label{fig:simulation}
\end{figure}

We set the initial $V_i(0)$ as orthogonal (and thus, independent) to the others, aiming to get insight into the situation in which the agent adjusts their differences from the most radical status. As more iteration $t$ passed by, genealogy vectors were synchronized according to the proximity matrix. In the real world, new concepts are consistently introduced to society, and thus the synchronization is hard to reach, yet we neglected the introduction of a new concept. As a result, the simulation ended with an identical vector after all. Our motivation was to obtain a general insight from various proximities. Hence, we captured the most optimized synchronization case by calculating the modeled similarity $S^{model} \left( t \right) \in \mathcal{R}^{N_{l}\times N_l}$ with a pairwise Euclidean distance with normalization for every iteration $t$ (see Methods, the distance is expressed by eq.~\ref{eq:distance_normalzation}). Then, we chose the final simulation result for the given proximity $S^{model}\left( t^* \right)$ as

  \begin{equation}
      	S^{model}\left( t^* \right) = \operatorname*{argmin}_t ||S^{model}\left( t \right) - S^{empirical}||_F,
  \end{equation}
  
\noindent where $S^{empirical}$ is the empirical knowledge structure similarity from Wikipedia’s knowledge network, and $|| \cdot ||_F$ is the Frobenius norm of a vector. For example, a Frobenius norm between the reconstructed similarity and actual similarity has a minimum value at $t=399$, as depicted in the snapshot of Facebook SCI proximity in Fig.~\ref{fig:simulation}b. Therefore, we choose $S^{model}\left( 399 \right)$ as the final state of the simulation. 

We test the model with the six empirical proximities (geographical distance, export, paper citations, paper collaboration, international student, and Facebook SCI), along with one random proximity as a null model. As the geographical distance is not a proximity measure, we use its log-transformed reciprocal as the geographical proximity. For all other cases, we use log-transformed proximity, similar to the empirical analysis (see Methods). As an illustrative example of our model results, we present the knowledge similarity matrix of the Facebook SCI from our model and empirical data in Fig.~\ref{fig:simulation}c, which shows similar structures.

Previously, we displayed the association between the knowledge similarity and the socio-economic proximities, which is high for the countries that exchange more. The results of our simple synchronization model are consistent with the empirical observations. We show a pairwise Pearson correlation $r$ between the similarity matrix of the model and empirical observation, as an indicator of how well the association pattern has been reproduced in our model. We found a similar increasing pattern of the association with our empirical observations above (Fig.~\ref{fig:simulation}d). Specifically, we observed the lowest Pearson correlation for the geographical distance ($r\simeq 0$), followed by the amount of export ($r=0.091$), the paper citation number ($r=0.174$), and the paper collaboration number ($r=0.225$). One exception was the order between the number of international students and Facebook SCI, which showed the third highest as well as the highest coefficient of determination respectively, in our empirical observations (Fig.~\ref{fig:regression}a). From our model, Facebook SCI showed a Pearson correlation $r=0.239$, while the international student proximity number showed a Pearson correlation of $r=0.269$. Note that the international student numbers are highly asymmetric originating, because it has been collected for only the inbound international students studying in OECD countries. To compensate for this asymmetry impact, we also tested the model with the symmetrized count of international students, by averaging the number of inbound and outbound students. The symmetrized model showed a significantly lower correlation than that of the asymmetry model ($r=0.216$), which is in-between the paper citation and the paper collaboration as similar as the empirical observation. Because of the knowledge structure synchronization through multiple channels, the current state of similarity is the result of an accumulated exchange process through many routes, which even include factors we have neglected. Our model is the simplest replica, using only a single route of exchange, but implies that social interaction can shape the structure of human knowledge. Moreover, the observed similarity was more robust for virtual connections, which may overcome geographical barriers in contemporary society. 
 
\section*{Discussion}
Humankind has accumulated and exchanged knowledge through various channels over time, facilitated by the technology of the time. Society has gradually progressed toward more efficient commutation, from physically proximate communication routes to virtual online interactions. In this study, we explore the similarity of knowledge structures between users of different languages.  We compare the similarities with socio-economic proximities, to identify the main route of contemporary knowledge exchange. Our results indicate the importance of both scientific and social connectedness in knowledge exchange, which shows the significant association between knowledge structures. Thus, this observation indicates that the changes in the main channel of knowledge exchange are from physical contact to online interactions. Our mechanistic model was motivated by the synchronization phenomena proposed to investigate the hidden mechanism behind the current state of knowledge similarity. The model replicated the interactions between language usage groups and reproduced the trend of knowledge similarities. Both the empirical data and model revealed a key factor of knowledge exchange: that is, socio-economic interaction led to a synchronization of the knowledge structures between different cultural areas.

Our approach has important implications for science studies, as online collaborative knowledge can provide non-experts with valuable insight into knowledge dissemination. This is a difficult assignment for traditional data set (e.g., papers or patents), which focus primarily on the knowledge structure of professionals, who make up just a small part of society overall. The restrictions on data collection, on the other hand, provide homework for future research. First, one may argue that Wikipedia’s accessibility (or inaccessibility in some regions) can affect results. The Chinese people living in mainland China have been unable to access Wikipedia since 2015 (it remains inaccessible). Nonetheless, the Chinese version of Wikipedia is one of the website’s most active language editions, and it plays an important role in the similarity network. These findings make it difficult to attribute knowledge structure to a specific geographical region. Second, OECD international student data did not include data for non-OECD countries. In addition to knowledge structure similarities and socio-economic data, this issue prevented direct comparison between the data and the model. Our knowledge structure was also focused on language users, rather than a certain country. As a result, we used language usage statistics to project languages onto countries. If a direct comparison were possible, it would provide additional information; however, we decided to leave this for future research.

We believe that Wikipedia data have considerable potential for future research. We investigated only the relationships between categories, pages, and language editions; nevertheless, there are billions of records with article content or user statistics that could be explored. Language link data and Wikidata’s curated collection provide well-structured, high-quality multilingual linkages that connect semantically similar objects. We show that language usage groups have diverse knowledge structures, indicating that, even if people face the same issue, they may have different perspectives on it. Quantifying the differences in interest changes based on their spoken language may be beneficial. Furthermore, unprecedented contemporary global problems, such as the COVID-19 pandemic, threaten to cause significant changes in how people work worldwide~\cite{brynjolfsson2020covid, yang2021effects} and collaborate~\cite{lee2021scientific}. Such changes may accelerate non-physical interactions for knowledge exchange. By pinpointing the primary paths of knowledge diffusion, we want to shed light on the unknown mechanism of general rules of knowledge evolution. Therefore, we would like to emphasize that our study is not simply restricted to Wikipedia, but has the potential for broader applications in contemporary society. 

\section*{Methods}
\subsection*{Description of Wikipedia data set}
We used Wikipedia SQL dump of $59$ different language editions on February 1, 2019. A list of language editions and their abbreviation are provided in Table.~\ref{supp:table:language_code}. Specifically, we used two collections of the Wikipedia dump: category membership link records (\texttt{*-categorylinks.sql.gz}) and interlanguage link records (\texttt{*-langlinks.sql.gz}). First, category membership link records contained directed linkage between a category and other items (e.g., page and category) in Wikipedia. We filtered \texttt{page} $\rightarrow$ \texttt{category} and \texttt{category} $\rightarrow$ \texttt{category} relationship (e.g., \texttt{page:Complex system} $\rightarrow$ \texttt{category:System theory}) to extract the reference relationship between scientific concepts. Second, interlanguage link records contained information of items in other language editions that were identical or reasonably similar to the source article. For instance, \texttt{page:Complex system} in English Wikipedia has a language linkage with \texttt{page:Syst\`eme complexe} in French Wikipedia, indicating that the two documents on this topic are identical.

\subsection*{Genealogy vector of scientific concept}
We calculated genealogy vectors of a given subject for each edition using our variant of personalized page rank (PPR) algorithm~\cite{jeh2003scaling}. The PPR is a node ranking algorithm with respect to a specific source node using the random walker on networks. For every timestep, the random walker moved to a nearby node chosen randomly with a probability proportionate to the edge weight between them, while the walker could return to the starting node, with a chance of fixed probability $\alpha$. Thus, the stationary distribution of the random walker starting from node $i$, denoted by $p_i=(p_{ik})$ was given by
  
  \begin{equation}
      	p_i = (1 - \alpha) W p_i +  \alpha v_i,
  \end{equation}

\noindent where $W$ is an adjacency matrix for a given network and $v_i$ is a column vector of length $N$ (number of nodes in the network) whose elements are zero except $i$th element equals to one. The teleport probability $\alpha$ is a tunable parameter, for which we used $\alpha = 0.3$ in this study.
  
For hierarchies of the structure of knowledge itself, we introduced a hierarchical bias on the transition matrix, $W$. First, we defined $l_i$ as the shortest path of node $i$ from the root node, practically interpreted as the level of node $i$. Therefore, a given starting node, $i$, transition matrix $W=(W_{ij})$ was defined as follows:
  
  \begin{equation}
      	W_{ij} = \frac{A_{ij}*k^{l_{i}}}{\sum_{j} {A_{ij}*k^{l_{j}}}},
  \end{equation}
 
\noindent where $A_{ij}$ is the adjacency matrix and $k$ is the tunable hyperparameter that controls the behavior of the random walker toward the hierarchy. If $k > 1$, the random walker was more likely to visit lower-level nodes, whereas the random walker tended to visit the higher-level nodes when $k < 1$. In this study, we used $k=0.5$ considering the hierarchy from the root node for a given subject. The PPR value of the source node is $p_{ii}\simeq\alpha$ by definition, although we forcibly assigned $p_{ii} = 0$ to remove the self-preference of the genealogy vectors. Then, we normalized genealogy vectors so that the sum of the vector is 1.

\subsection*{Calculation of subject similarity and knowledge structure similarity}

We hypothesized that the interlanguage similarities between genealogy vectors of the same subject could be used as cognitive similarities between them. We presented a simple example to depict the computation of similarity when one subject is solely connected to another subject, and we present more complex cases (\textit{e.g.} many-to-many) in Supplementary Information (Fig.~\ref{suppfig:synoym}). For the most straightforward and common case, we first defined this similarity as follows:

  \begin{equation}
      	d_{x}^{a \rightarrow b} = d(p_x^{a} T^{a \rightarrow b}, p_x^{b}),
  \end{equation}

\noindent where $p_x^{a}$ and $p_x^{b}$ are genealogy vectors of subject $x$ in the knowledge network of $a$ language edition and $b$ language edition, respectively, and $T^{a \rightarrow b}$ is translation matrix between two different knowledge network from language link data where $T^{a \rightarrow b}_{ij} = 1$ if there is link between subject $i$ in the language $a$ and $j$ subject $j$ in the language $b$. We set $T^{a \rightarrow b}_{ij} = 0$ otherwise. For the distance function $d$, we used the $l_2$ Euclidean distance.

We then defined subject similarities between the same subject in different language editions as

  \begin{equation}
  s_{x}^{a \rightarrow b} = \frac{\sqrt{2} - d_{x}^{\alpha \rightarrow \beta}}{\sqrt{2}},
  \label{eq:distance_normalzation}
  \end{equation}

\noindent where $\sqrt{2}$ is the theoretical maximum value of the distance between vectors whose elements are positive, and their sum is one. Finally, we defined the knowledge structure similarity by aggregating the similarities between two language editions by averaging the similarity of all co-existing concepts as follows:

  \begin{equation}
  s^{a \rightarrow b} = \frac{\sum_{x \in \mathcal {A}}{s_{x}^{a \rightarrow b}}}{|\mathcal {A}|},
  \end{equation}

\noindent where $\mathcal{A}$ is the subject that co-exists in both language editions.

\subsection*{Community detection}
We employed Leiden algorithms \cite{traag2019louvain} to find community structure from the knowledge structure similarity network, which is a refined version of the well-known Louvain algorithm. We used a quality function $Q$ of the Potts model with the configuration null mode \cite{leicht2008community} as follows:

  \begin{equation}
    Q = \sum_{ij} 	\left( A_{ij} - \gamma \frac{k_i^{out}k_j^{in}}{m} \right)\delta(\sigma_i, \sigma_j),
  \end{equation}
  
\noindent where $k_i^{out}$ and $k_i^{in}$ are the out-strength and in-strength of node $i$, respectively, $A$ is the weighted adjacency matrix, $m$ is the total edge weight, and $\sigma_i$ denotes the membership of node $i$. Here, $\gamma$ is the resolution parameter, where $\delta(\sigma_i, \sigma_j)=1$ if $\sigma_i =\sigma_j$ and 0 otherwise. We may control resolution parameter $\gamma$ to vary the number of clusters, and we use $\gamma$ with a default value of 1. A modularity value of the obtained community is 167.29.

\subsection*{Description of socio-economic data sets}
To verify our hypothesis that social connection yields the similarity in knowledge structure, we collected additional country-to-country socio-economic datasets. Although we collected 59 language editions, only 52 languages exist in the country-to-country socio-economic data because it is difficult to find the usage statistics for languages such as Bosnian, Welsh, Croatian, Norwegian Nynorsk, Scottish, Serbian, and Cantonese. The export data were extracted from two different sources: IMF Data in December 2019 and UN Comtrade export data in January 2020. We obtained the statistics of scientific papers (citations and collaborations) from SCOPUS’s April 2019 data; patent information was retrieved from PATSTAT’s Spring 2019 data. The international student count was collected from OECD in December 2019. Please note that only inbound international student numbers in OECD countries were collected, and thus, the statistics are highly asymmetric and incomplete because it does not provide a number for international students in non-OECD countries. Facebook Social Connected Index (SCI) is the index indicating the degree of the social connection between the two regions, which has been used in various disciplines recently~\cite{bailey2018social, bailey2020social, vahedi2021predicting, du2021international}. All socio-economic data was directed, except the paper/patent collaboration and Facebook SCI.
 
\subsection*{Mapping the country-level statistics onto the language}
The socio-economic data described earlier were county-level data, whereas our similarity measure was language-level statistics. For our analysis, we projected country-to-country data to language-to-language data using the language profile for each country. 

Consider a country-level-statistics, $X \in \mathcal{R}^{N_c * N_c}$, where $N_c$ is the total number of unique countries in the dataset and $X_{ij}$ denotes the socio-economic quantity between countries $i$ and $j$. To map this matrix onto the language space, we constructed a country to a language projection matrix, $A \in \mathcal{R}^{N_c * N_l}$, where $N_l$ is the total number of unique languages. The elements of matrix $A$ were obtained from the language usage profile of countries. We assigned the proportion of language $a$ in country $i$ onto $A_{i\alpha}$. For instance, $A_{\text{English, United States}}$ was 0.821 because the usage share of English in the United States is 82.1\%. Using this metric, we constructed the language-to-language socio-economic data, $Y$, with simple projection $Y = A^T X A$, where $Y_{a b}$ means projected socio-economic quantity between languages $a$ and $b$.

\subsection*{Initialization of the genealogy vectors for the model simulation.}
For the pair model study, we initialized vectors as orthogonal to each other. We assigned the dimension of the vectors to be a multiple of the number of artificial user groups. Otherwise, the simulation result was biased toward a set of vectors that were not orthogonal or had more nonzero elements. Accordingly, we set each row with an equal number of equally weighted nonzero values (\textit{e.g.}, $1$). From the orthogonal condition, each column had only one nonzero value so that $Rank(V)$ was equal to the number of user groups. Then, we normalized each row similar to the empirical genealogy vectors. We simulated the model with 52 language usage groups and using 520 dimensions, resulting genealogy matrix $V=\mathcal{R}^{52 \times 520}$ for the results in the main text. 

\section*{Acknowledgements}
We thank M. Ahn, I. Hong, H. Kim, L. Miao, and Y.-Y. Ahn for their helpful discussions. This work was supported by the National Research Foundation of Korea (NRF) with grant number NRF-2021R1F1A106303011 (J.Y.; W.S.J.) and NRF-2020R1A2C1100489 (J.Y.). The Korea Institute of Science and Technology Information (KISTI) also offered institutional support for this work (K-22-L03-C01; J.P.) and provided KREONET, our high-speed internet connection. We would also like to thank Facebook Inc., for making the Social Connectedness Index dataset available to us.

\section*{Author Contributions}
All authors contributed to the work presented in this paper. Jisung Yoon was involved in conceptualization, analysis, and writing. Jinseo Park contributed to data collecting and writing. Woo-sung Jung and Jinhyuk Yun contributed to conceptualization and writing. All authors discussed the results and commented on the manuscript at all stages.

\section*{Additional Information}
Supplementary Information is available for this paper. Correspondence and requests for materials should be addressed to Dr. Jinhuk Yun and Dr. Woo-Sung Jung.

\section*{Data availability}
Wikipedia data are available at wiki-dumps, \url{https://dumps.wikimedia.org/}, export data are available at \url{https://data.imf.org/?sk=9d6028d4-f14a-464c-a2f2-59b2cd424b85} (IMF) and \url{https://comtrade.un.org/} (UN), and Facebook Social Connected Index is available at \url{https://dataforgood.facebook.com/dfg/tools/social-connectedness-index}. Paper (SCOPUS) and patent (PATSTAT) data can be accessed under a license agreement, which cannot share publicly due to the data’s copyright.
 
 \section*{Code Availability}
The code used in this analysis can be found at \url{https://github.com/jisungyoon/Structure-of-Science}.

\bibliographystyle{naturemag}
\bibliography{main}

\begin{thebibliography}{10}
\expandafter\ifx\csname url\endcsname\relax
  \def\url#1{\texttt{#1}}\fi
\expandafter\ifx\csname urlprefix\endcsname\relax\def\urlprefix{URL }\fi
\providecommand{\bibinfo}[2]{#2}
\providecommand{\eprint}[2][]{\url{#2}}

\bibitem{code1980language}
\bibinfo{author}{Code, L.}
\newblock \bibinfo{title}{Language and knowledge}.
\newblock \emph{\bibinfo{journal}{Word}} \textbf{\bibinfo{volume}{31}},
  \bibinfo{pages}{245--258} (\bibinfo{year}{1980}).

\bibitem{grimm2014understanding}
\bibinfo{author}{Grimm, S.~R.}
\newblock \bibinfo{title}{Understanding as knowledge of causes}.
\newblock In \emph{\bibinfo{booktitle}{Virtue epistemology naturalized}},
  \bibinfo{pages}{329--345} (\bibinfo{publisher}{Springer},
  \bibinfo{year}{2014}).

\bibitem{kant2000critique}
\bibinfo{author}{Kant, I.}
\newblock \emph{\bibinfo{title}{Critique of the Power of Judgment}}
  (\bibinfo{publisher}{Cambridge University Press}, \bibinfo{year}{2000}).

\bibitem{schieffelin1986language}
\bibinfo{author}{Schieffelin, B.~B.} \& \bibinfo{author}{Ochs, E.}
\newblock \bibinfo{title}{Language socialization}.
\newblock \emph{\bibinfo{journal}{Annual Review of Anthropology}}
  \textbf{\bibinfo{volume}{15}}, \bibinfo{pages}{163--191}
  (\bibinfo{year}{1986}).

\bibitem{andrea2014silk}
\bibinfo{author}{Andrea, A.~J.}
\newblock \bibinfo{title}{The silk road in world history: A review essay}.
\newblock \emph{\bibinfo{journal}{Asian Review of World Histories}}
  \textbf{\bibinfo{volume}{2}}, \bibinfo{pages}{105--127}
  (\bibinfo{year}{2014}).

\bibitem{lu2016earliest}
\bibinfo{author}{Lu, H.} \emph{et~al.}
\newblock \bibinfo{title}{Earliest tea as evidence for one branch of the silk
  road across the tibetan plateau}.
\newblock \emph{\bibinfo{journal}{Scientific Reports}}
  \textbf{\bibinfo{volume}{6}}, \bibinfo{pages}{1--8} (\bibinfo{year}{2016}).

\bibitem{bhandari2011global}
\bibinfo{author}{Bhandari, R.} \& \bibinfo{author}{Blumenthal, P.}
\newblock \bibinfo{title}{Global student mobility and the twenty-first century
  silk road: National trends and new directions}.
\newblock In \emph{\bibinfo{booktitle}{International students and global
  mobility in higher education}}, \bibinfo{pages}{1--23}
  (\bibinfo{publisher}{Springer}, \bibinfo{year}{2011}).

\bibitem{inkpen2005social}
\bibinfo{author}{Inkpen, A.~C.} \& \bibinfo{author}{Tsang, E.~W.}
\newblock \bibinfo{title}{Social capital, networks, and knowledge transfer}.
\newblock \emph{\bibinfo{journal}{Academy of management review}}
  \textbf{\bibinfo{volume}{30}}, \bibinfo{pages}{146--165}
  (\bibinfo{year}{2005}).

\bibitem{wu2007fostering}
\bibinfo{author}{Wu, W.-L.}, \bibinfo{author}{Hsu, B.-F.} \&
  \bibinfo{author}{Yeh, R.-S.}
\newblock \bibinfo{title}{Fostering the determinants of knowledge transfer: a
  team-level analysis}.
\newblock \emph{\bibinfo{journal}{Journal of Information Science}}
  \textbf{\bibinfo{volume}{33}}, \bibinfo{pages}{326--339}
  (\bibinfo{year}{2007}).

\bibitem{ringberg2008towards}
\bibinfo{author}{Ringberg, T.} \& \bibinfo{author}{Reihlen, M.}
\newblock \bibinfo{title}{Towards a socio-cognitive approach to knowledge
  transfer}.
\newblock \emph{\bibinfo{journal}{Journal of Management Studies}}
  \textbf{\bibinfo{volume}{45}}, \bibinfo{pages}{912--935}
  (\bibinfo{year}{2008}).

\bibitem{welch2008importance}
\bibinfo{author}{Welch, D.~E.} \& \bibinfo{author}{Welch, L.~S.}
\newblock \bibinfo{title}{The importance of language in international knowledge
  transfer}.
\newblock \emph{\bibinfo{journal}{Management International Review}}
  \textbf{\bibinfo{volume}{48}}, \bibinfo{pages}{339--360}
  (\bibinfo{year}{2008}).

\bibitem{ambos2009impact}
\bibinfo{author}{Ambos, T.~C.} \& \bibinfo{author}{Ambos, B.}
\newblock \bibinfo{title}{The impact of distance on knowledge transfer
  effectiveness in multinational corporations}.
\newblock \emph{\bibinfo{journal}{Journal of International Management}}
  \textbf{\bibinfo{volume}{15}}, \bibinfo{pages}{1--14} (\bibinfo{year}{2009}).

\bibitem{qian2009knowledge}
\bibinfo{author}{Qian, Y.}, \bibinfo{author}{Liang, J.} \&
  \bibinfo{author}{Dang, C.}
\newblock \bibinfo{title}{Knowledge structure, knowledge granulation and
  knowledge distance in a knowledge base}.
\newblock \emph{\bibinfo{journal}{International Journal of Approximate
  Reasoning}} \textbf{\bibinfo{volume}{50}}, \bibinfo{pages}{174--188}
  (\bibinfo{year}{2009}).

\bibitem{song2013detecting}
\bibinfo{author}{Song, M.} \& \bibinfo{author}{Kim, S.~Y.}
\newblock \bibinfo{title}{Detecting the knowledge structure of bioinformatics
  by mining full-text collections}.
\newblock \emph{\bibinfo{journal}{Scientometrics}}
  \textbf{\bibinfo{volume}{96}}, \bibinfo{pages}{183--201}
  (\bibinfo{year}{2013}).

\bibitem{su2010mapping}
\bibinfo{author}{Su, H.-N.} \& \bibinfo{author}{Lee, P.-C.}
\newblock \bibinfo{title}{Mapping knowledge structure by keyword co-occurrence:
  a first look at journal papers in technology foresight}.
\newblock \emph{\bibinfo{journal}{Scientometrics}}
  \textbf{\bibinfo{volume}{85}}, \bibinfo{pages}{65--79}
  (\bibinfo{year}{2010}).

\bibitem{hu2014empirical}
\bibinfo{author}{Hu, Z.}, \bibinfo{author}{Fang, S.} \& \bibinfo{author}{Liang,
  T.}
\newblock \bibinfo{title}{Empirical study of constructing a knowledge
  organization system of patent documents using topic modeling}.
\newblock \emph{\bibinfo{journal}{Scientometrics}}
  \textbf{\bibinfo{volume}{100}}, \bibinfo{pages}{787--799}
  (\bibinfo{year}{2014}).

\bibitem{sakata2012identifying}
\bibinfo{author}{Sakata, I.}, \bibinfo{author}{Sasaki, H.} \&
  \bibinfo{author}{Kajikawa, Y.}
\newblock \bibinfo{title}{Identifying knowledge structure of patent and
  innovation research}.
\newblock \emph{\bibinfo{journal}{Journal of Intellectual Property Association
  of Japan}} \textbf{\bibinfo{volume}{8}}, \bibinfo{pages}{56--67}
  (\bibinfo{year}{2012}).

\bibitem{fortunato2018science}
\bibinfo{author}{Fortunato, S.} \emph{et~al.}
\newblock \bibinfo{title}{Science of science}.
\newblock \emph{\bibinfo{journal}{Science}} \textbf{\bibinfo{volume}{359}}
  (\bibinfo{year}{2018}).

\bibitem{yasseri2012circadian}
\bibinfo{author}{Yasseri, T.}, \bibinfo{author}{Sumi, R.} \&
  \bibinfo{author}{Kert{\'e}sz, J.}
\newblock \bibinfo{title}{Circadian patterns of wikipedia editorial activity: A
  demographic analysis}.
\newblock \emph{\bibinfo{journal}{PloS one}} \textbf{\bibinfo{volume}{7}},
  \bibinfo{pages}{e30091} (\bibinfo{year}{2012}).

\bibitem{yasseri2012dynamics}
\bibinfo{author}{Yasseri, T.}, \bibinfo{author}{Sumi, R.},
  \bibinfo{author}{Rung, A.}, \bibinfo{author}{Kornai, A.} \&
  \bibinfo{author}{Kert{\'e}sz, J.}
\newblock \bibinfo{title}{Dynamics of conflicts in wikipedia}.
\newblock \emph{\bibinfo{journal}{PloS one}} \textbf{\bibinfo{volume}{7}},
  \bibinfo{pages}{e38869} (\bibinfo{year}{2012}).

\bibitem{yun2019early}
\bibinfo{author}{Yun, J.}, \bibinfo{author}{Lee, S.~H.} \&
  \bibinfo{author}{Jeong, H.}
\newblock \bibinfo{title}{Early onset of structural inequality in the formation
  of collaborative knowledge in all wikimedia projects}.
\newblock \emph{\bibinfo{journal}{Nature human behaviour}}
  \textbf{\bibinfo{volume}{3}}, \bibinfo{pages}{155--163}
  (\bibinfo{year}{2019}).

\bibitem{samoilenko2016linguistic}
\bibinfo{author}{Samoilenko, A.}, \bibinfo{author}{Karimi, F.},
  \bibinfo{author}{Edler, D.}, \bibinfo{author}{Kunegis, J.} \&
  \bibinfo{author}{Strohmaier, M.}
\newblock \bibinfo{title}{Linguistic neighbourhoods: explaining cultural
  borders on wikipedia through multilingual co-editing activity}.
\newblock \emph{\bibinfo{journal}{EPJ data science}}
  \textbf{\bibinfo{volume}{5}}, \bibinfo{pages}{1--20} (\bibinfo{year}{2016}).

\bibitem{karimi2015mapping}
\bibinfo{author}{Karimi, F.}, \bibinfo{author}{Bohlin, L.},
  \bibinfo{author}{Samoilenko, A.}, \bibinfo{author}{Rosvall, M.} \&
  \bibinfo{author}{Lancichinetti, A.}
\newblock \bibinfo{title}{Mapping bilateral information interests using the
  activity of wikipedia editors}.
\newblock \emph{\bibinfo{journal}{Palgrave Communications}}
  \textbf{\bibinfo{volume}{1}}, \bibinfo{pages}{1--7} (\bibinfo{year}{2015}).

\bibitem{el2018capturing}
\bibinfo{author}{El~Zant, S.}, \bibinfo{author}{Jaffr{\`e}s-Runser, K.} \&
  \bibinfo{author}{Shepelyansky, D.~L.}
\newblock \bibinfo{title}{Capturing the influence of geopolitical ties from
  wikipedia with reduced google matrix}.
\newblock \emph{\bibinfo{journal}{Plos one}} \textbf{\bibinfo{volume}{13}},
  \bibinfo{pages}{e0201397} (\bibinfo{year}{2018}).

\bibitem{zesch2007analysis}
\bibinfo{author}{Zesch, T.} \& \bibinfo{author}{Gurevych, I.}
\newblock \bibinfo{title}{Analysis of the wikipedia category graph for nlp
  applications}.
\newblock In \emph{\bibinfo{booktitle}{Proceedings of the Second Workshop on
  TextGraphs: Graph-Based Algorithms for Natural Language Processing}},
  \bibinfo{pages}{1--8} (\bibinfo{year}{2007}).

\bibitem{nastase2008decoding}
\bibinfo{author}{Nastase, V.} \& \bibinfo{author}{Strube, M.}
\newblock \bibinfo{title}{Decoding wikipedia categories for knowledge
  acquisition.}
\newblock In \emph{\bibinfo{booktitle}{AAAI}}, vol.~\bibinfo{volume}{8},
  \bibinfo{pages}{1219--1224} (\bibinfo{year}{2008}).

\bibitem{schonhofen2009identifying}
\bibinfo{author}{Sch{\"o}nhofen, P.}
\newblock \bibinfo{title}{Identifying document topics using the wikipedia
  category network}.
\newblock \emph{\bibinfo{journal}{Web Intelligence and Agent Systems: An
  International Journal}} \textbf{\bibinfo{volume}{7}},
  \bibinfo{pages}{195--207} (\bibinfo{year}{2009}).

\bibitem{ponzetto2009large}
\bibinfo{author}{Ponzetto, S.~P.} \& \bibinfo{author}{Navigli, R.}
\newblock \bibinfo{title}{Large-scale taxonomy mapping for restructuring and
  integrating wikipedia}.
\newblock In \emph{\bibinfo{booktitle}{Twenty-First International Joint
  Conference on Artificial Intelligence}} (\bibinfo{year}{2009}).

\bibitem{yoon_build_2018}
\bibinfo{author}{Yoon, J.}, \bibinfo{author}{Yun*, J.} \&
  \bibinfo{author}{Jung*, W.-S.}
\newblock \bibinfo{title}{Build {Up} of a {Subject} {Classification} {System}
  from {Collective} {Intelligence}}.
\newblock \emph{\bibinfo{journal}{New Physics: Sae Mulli}}
  \textbf{\bibinfo{volume}{68}}, \bibinfo{pages}{647--654}
  (\bibinfo{year}{2018}).
\newblock
  \urlprefix\url{http://www.npsm-kps.org/journal/DOIx.php?id=10.3938/NPSM.68.647}.

\bibitem{jeh2003scaling}
\bibinfo{author}{Jeh, G.} \& \bibinfo{author}{Widom, J.}
\newblock \bibinfo{title}{Scaling personalized web search}.
\newblock In \emph{\bibinfo{booktitle}{Proceedings of the 12th International
  Conference on World Wide Web}}, \bibinfo{pages}{271--279}
  (\bibinfo{year}{2003}).

\bibitem{kuramoto1975international}
\bibinfo{author}{Kuramoto, Y.}
\newblock \bibinfo{title}{International symposium on mathematical problems in
  theoretical physics}.
\newblock \emph{\bibinfo{journal}{Lecture notes in Physics}}
  \textbf{\bibinfo{volume}{30}}, \bibinfo{pages}{420} (\bibinfo{year}{1975}).

\bibitem{waltman2020principled}
\bibinfo{author}{Waltman, L.}, \bibinfo{author}{Boyack, K.~W.},
  \bibinfo{author}{Colavizza, G.} \& \bibinfo{author}{van Eck, N.~J.}
\newblock \bibinfo{title}{A principled methodology for comparing relatedness
  measures for clustering publications}.
\newblock \emph{\bibinfo{journal}{Quantitative Science Studies}}
  \textbf{\bibinfo{volume}{1}}, \bibinfo{pages}{691--713}
  (\bibinfo{year}{2020}).

\bibitem{traag2019louvain}
\bibinfo{author}{Traag, V.~A.}, \bibinfo{author}{Waltman, L.} \&
  \bibinfo{author}{Van~Eck, N.~J.}
\newblock \bibinfo{title}{From louvain to leiden: guaranteeing well-connected
  communities}.
\newblock \emph{\bibinfo{journal}{Scientific reports}}
  \textbf{\bibinfo{volume}{9}}, \bibinfo{pages}{1--12} (\bibinfo{year}{2019}).

\bibitem{ronen2014links}
\bibinfo{author}{Ronen, S.} \emph{et~al.}
\newblock \bibinfo{title}{Links that speak: The global language network and its
  association with global fame}.
\newblock \emph{\bibinfo{journal}{Proceedings of the National Academy of
  Sciences}} \textbf{\bibinfo{volume}{111}}, \bibinfo{pages}{E5616--E5622}
  (\bibinfo{year}{2014}).

\bibitem{pithouse2009making}
\bibinfo{author}{Pithouse, K.}, \bibinfo{author}{Mitchell, C.} \&
  \bibinfo{author}{Moletsane, R.}
\newblock \emph{\bibinfo{title}{Making connections: Self-study \& social
  action}}, vol. \bibinfo{volume}{357} (\bibinfo{publisher}{Peter Lang},
  \bibinfo{year}{2009}).

\bibitem{heese1971herkoms}
\bibinfo{author}{Heese, J.~A.}
\newblock \emph{\bibinfo{title}{Die herkoms van die Afrikaner, 1657-1867}}
  (\bibinfo{year}{1971}).

\bibitem{hummels2007transportation}
\bibinfo{author}{Hummels, D.}
\newblock \bibinfo{title}{Transportation costs and international trade in the
  second era of globalization}.
\newblock \emph{\bibinfo{journal}{Journal of Economic perspectives}}
  \textbf{\bibinfo{volume}{21}}, \bibinfo{pages}{131--154}
  (\bibinfo{year}{2007}).

\bibitem{ethnologue}
\bibinfo{title}{Ethnologue global dataset}.
\newblock \bibinfo{howpublished}{\url{https://www.ethnologue.com/}}.
\newblock \bibinfo{note}{Accessed: 2019-10-30}.

\bibitem{murray2020unsupervised}
\bibinfo{author}{Murray, D.} \emph{et~al.}
\newblock \bibinfo{title}{Unsupervised embedding of trajectories captures the
  latent structure of mobility}.
\newblock \emph{\bibinfo{journal}{arXiv preprint arXiv:2012.02785}}
  (\bibinfo{year}{2020}).

\bibitem{OECD2011}
\bibinfo{author}{OECD}.
\newblock \emph{\bibinfo{title}{How many international students stay on in the
  host country?}} (\bibinfo{year}{2011}).
\newblock
  \urlprefix\url{https://www.oecd-ilibrary.org/content/component/eag_highlights-2011-14-en}.

\bibitem{gueguen2011similarity}
\bibinfo{author}{Gu{\'e}guen, N.}, \bibinfo{author}{Martin, A.} \&
  \bibinfo{author}{Meineri, S.}
\newblock \bibinfo{title}{Similarity and social interaction: When similarity
  fosters implicit behavior toward a stranger}.
\newblock \emph{\bibinfo{journal}{The Journal of social psychology}}
  \textbf{\bibinfo{volume}{151}}, \bibinfo{pages}{671--673}
  (\bibinfo{year}{2011}).

\bibitem{faraj2001web}
\bibinfo{author}{Faraj, S.} \& \bibinfo{author}{Wasko, M.~M.}
\newblock \bibinfo{title}{The web of knowledge: An investigation of knowledge
  exchange in networks of practice}.
\newblock \emph{\bibinfo{journal}{Paper submitted for publication}}
  (\bibinfo{year}{2001}).

\bibitem{peng2021neural}
\bibinfo{author}{Peng, H.}, \bibinfo{author}{Ke, Q.}, \bibinfo{author}{Budak,
  C.}, \bibinfo{author}{Romero, D.~M.} \& \bibinfo{author}{Ahn, Y.-Y.}
\newblock \bibinfo{title}{Neural embeddings of scholarly periodicals reveal
  complex disciplinary organizations}.
\newblock \emph{\bibinfo{journal}{Science Advances}}
  \textbf{\bibinfo{volume}{7}}, \bibinfo{pages}{eabb9004}
  (\bibinfo{year}{2021}).

\bibitem{brynjolfsson2020covid}
\bibinfo{author}{Brynjolfsson, E.} \emph{et~al.}
\newblock \bibinfo{title}{Covid-19 and remote work: an early look at us data}.
\newblock \bibinfo{type}{Tech. Rep.}, \bibinfo{institution}{National Bureau of
  Economic Research} (\bibinfo{year}{2020}).

\bibitem{yang2021effects}
\bibinfo{author}{Yang, L.} \emph{et~al.}
\newblock \bibinfo{title}{The effects of remote work on collaboration among
  information workers}.
\newblock \emph{\bibinfo{journal}{Nature human behaviour}}
  \bibinfo{pages}{1--12} (\bibinfo{year}{2021}).

\bibitem{lee2021scientific}
\bibinfo{author}{Lee, J.~J.} \& \bibinfo{author}{Haupt, J.~P.}
\newblock \bibinfo{title}{Scientific collaboration on covid-19 amidst
  geopolitical tensions between the us and china}.
\newblock \emph{\bibinfo{journal}{The Journal of Higher Education}}
  \textbf{\bibinfo{volume}{92}}, \bibinfo{pages}{303--329}
  (\bibinfo{year}{2021}).

\bibitem{leicht2008community}
\bibinfo{author}{Leicht, E.~A.} \& \bibinfo{author}{Newman, M.~E.}
\newblock \bibinfo{title}{Community structure in directed networks}.
\newblock \emph{\bibinfo{journal}{Physical review letters}}
  \textbf{\bibinfo{volume}{100}}, \bibinfo{pages}{118703}
  (\bibinfo{year}{2008}).

\bibitem{bailey2018social}
\bibinfo{author}{Bailey, M.}, \bibinfo{author}{Cao, R.},
  \bibinfo{author}{Kuchler, T.}, \bibinfo{author}{Stroebel, J.} \&
  \bibinfo{author}{Wong, A.}
\newblock \bibinfo{title}{Social connectedness: Measurement, determinants, and
  effects}.
\newblock \emph{\bibinfo{journal}{Journal of Economic Perspectives}}
  \textbf{\bibinfo{volume}{32}}, \bibinfo{pages}{259--80}
  (\bibinfo{year}{2018}).

\bibitem{bailey2020social}
\bibinfo{author}{Bailey, M.}, \bibinfo{author}{Farrell, P.},
  \bibinfo{author}{Kuchler, T.} \& \bibinfo{author}{Stroebel, J.}
\newblock \bibinfo{title}{Social connectedness in urban areas}.
\newblock \emph{\bibinfo{journal}{Journal of Urban Economics}}
  \textbf{\bibinfo{volume}{118}}, \bibinfo{pages}{103264}
  (\bibinfo{year}{2020}).

\bibitem{vahedi2021predicting}
\bibinfo{author}{Vahedi, B.}, \bibinfo{author}{Karimzadeh, M.} \&
  \bibinfo{author}{Zoraghein, H.}
\newblock \bibinfo{title}{Predicting county-level covid-19 cases using
  spatiotemporal machine learning: Modeling human interactions using social
  media and cell-phone data}  (\bibinfo{year}{2021}).

\bibitem{du2021international}
\bibinfo{author}{Du, Z.} \emph{et~al.}
\newblock \bibinfo{title}{International risk of the new variant covid-19
  importations originating in the united kingdom}.
\newblock \emph{\bibinfo{journal}{MedRxiv}}  (\bibinfo{year}{2021}).

\bibitem{ester1996density}
\bibinfo{author}{Ester, M.}, \bibinfo{author}{Kriegel, H.-P.},
  \bibinfo{author}{Sander, J.}, \bibinfo{author}{Xu, X.} \emph{et~al.}
\newblock \bibinfo{title}{A density-based algorithm for discovering clusters in
  large spatial databases with noise.}
\newblock In \emph{\bibinfo{booktitle}{kdd}}, vol.~\bibinfo{volume}{96},
  \bibinfo{pages}{226--231} (\bibinfo{year}{1996}).

\end{thebibliography}

\pagebreak
\begin{center}
\textbf{\Huge Supplementary Information:Quantifying knowledge synchronisation in the 21st century}
\end{center}
\setcounter{equation}{0}
\setcounter{figure}{0}
\setcounter{table}{0}
\makeatletter
\renewcommand{\theequation}{S\arabic{equation}}
\renewcommand{\thefigure}{S\arabic{figure}}

\begin{figure}[ht!]
    \centering
    \includegraphics[width=0.95\textwidth]{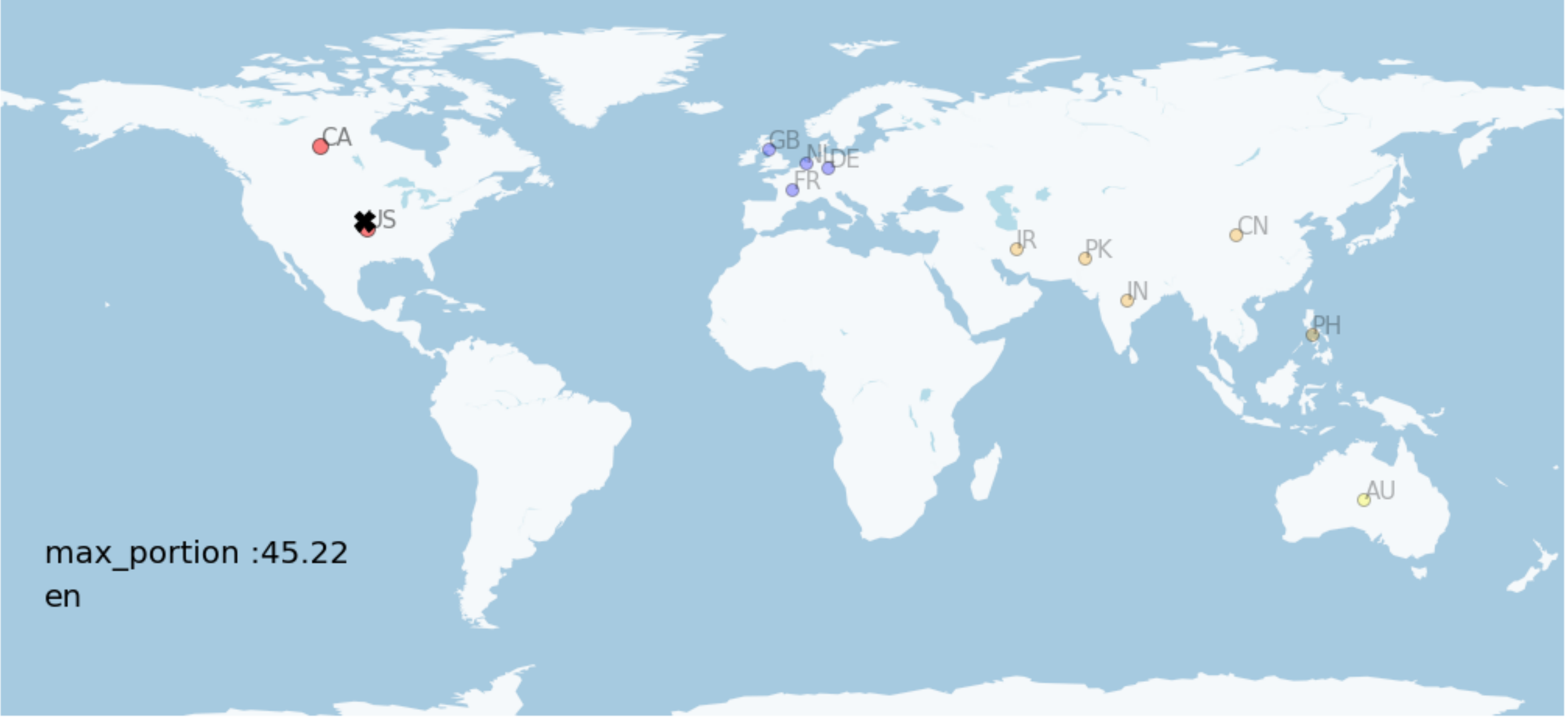}
    \caption{\textbf{Geographic location of the English} Four clusters are identified by using the DBSCAN method, and red dots represent countries belonging to the max. portion cluster. The black cross indicates the centroid of English.}
    \label{suppfig:locaiton_en}
\end{figure}

\begin{figure}[ht!]
    \centering
    \includegraphics[width=0.95\textwidth]{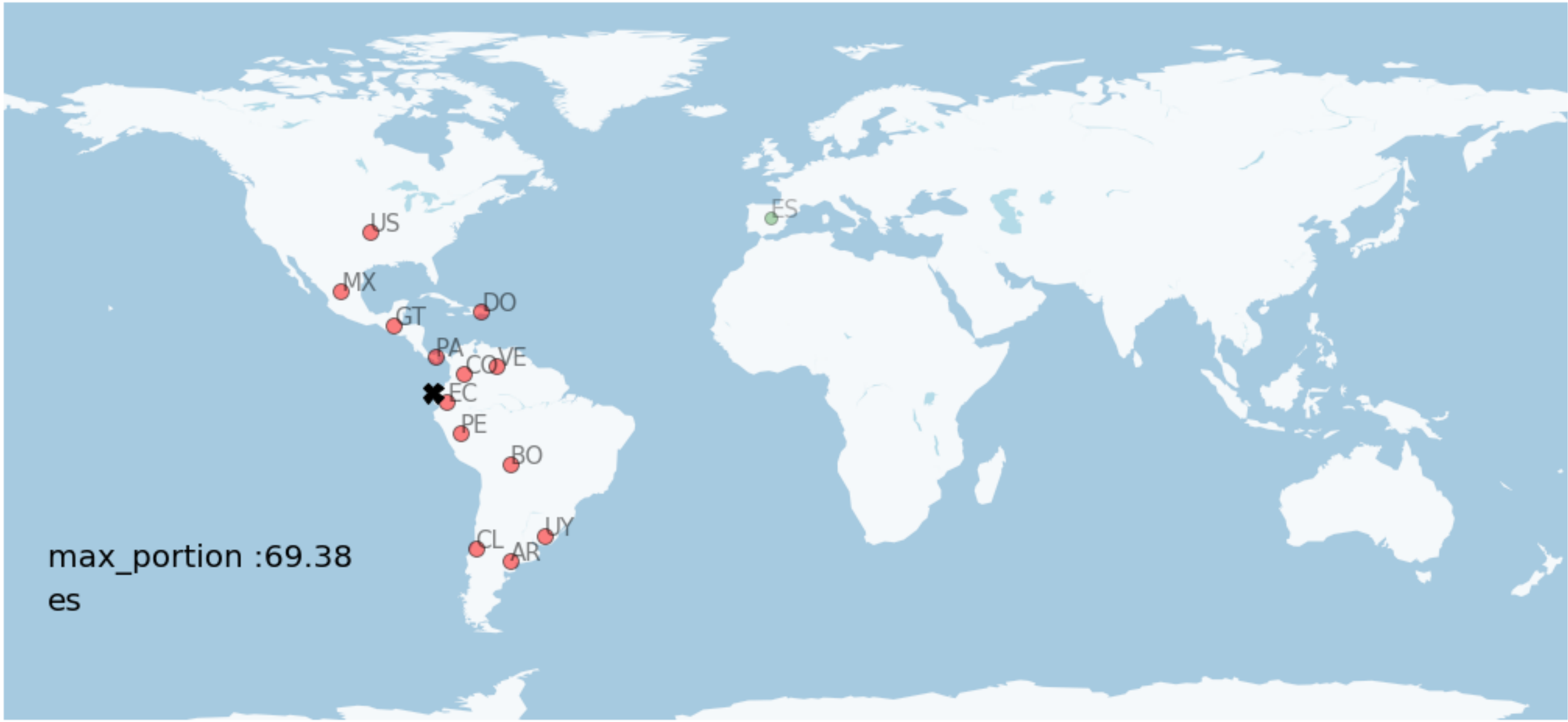}
    \caption{\textbf{Geographic location of the Spanish} Another example. Two clusters are identified by employing the DBSCAN method, and red dots represent countries belonging to the max. portion cluster. The black cross indicates the centroid of Spanish.}
    \label{suppfig:locaiton_es}
\end{figure}

\begin{figure}[ht!]
    \centering
    \includegraphics[width=0.95\textwidth]{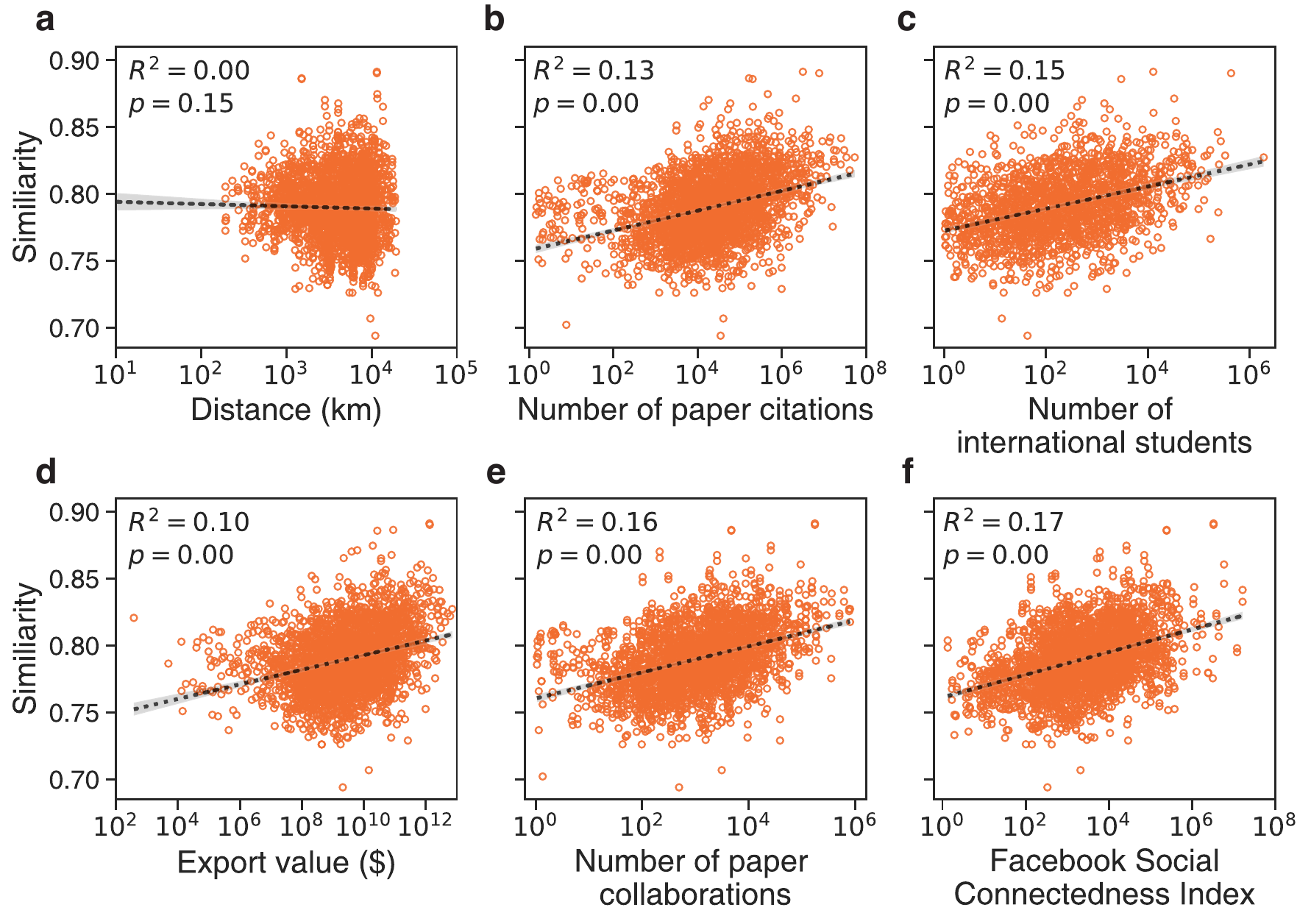}
    \caption{\textbf{Interrelationship of knowledge structures across language groups reveals the effect of socio-economic interactions with language usage data.} Compared with Fig.~\ref{fig:regression}, we calculate the centroid of the language group with language usage statistics~\cite{ethnologue}. The correlation between structural similarity of knowledge and socio-economic factors: \textbf{a.} geographical distance for the centroids of language pairs, \textbf{b.} Amount of exported goods for language pairs (IMF), \textbf{c.} Number of citations in papers for language pairs (SCOPUS), \textbf{d.} Number of co-authorships in paper for language pairs (SCOPUS), \textbf{e.} Number of the international students for language pairs (OECD), and \textbf{f.} Facebook Social Connected Index, the strength of connectedness between areas by represented by Facebook friendship ties, for language pairs.}
    \label{suppfig:regression_usage}
\end{figure}

\paragraph*{S1 Text. Geographic location of a language}
\label{si:text:geo-location of the language}
We determine geographic locations for languages using pageview by country statistics provided by Wikimedia Statistics \url{https://stats.wikimedia.org/}. Because visitors can be from anywhere in the world, page view data contain visit history from multiple countries. We sought to find the interrelation between geographical distance and knowledge similarity and obtained a single point as follows.
First, we used geographic information to determine the centroid of each country. We then conducted geo-location clustering by using a Density-Based Spatial Clustering of Applications with Noise (DBSCAN) method \cite{ester1996density} to extract the maximum portion cluster, to obtain the location in which the given language is mostly spoken. For example, as depicted in Fig.~\ref{suppfig:locaiton_en}, there are four clusters: North America (including Canada and the United States), Western Europe (including Great Britain), Oceania (including Australia), and the others (including Philippines and India). Among these clusters, the largest possessed one is North America, with the proportion of $45.22\%$. We considered the geographic location of a language to be the centroid of the largest posed cluster (\textit{e.g.} black cross in Fig.~\ref{suppfig:locaiton_en}). We present another example of Spanish in Fig.~\ref{suppfig:locaiton_es}. To confirm the robustness of our finding, we tested the alternative geographic location for a language using language usage statistics provided by Ethnologue \cite{ethnologue} and found that there was no significant difference between the choice of centroids (Fig.~\ref{suppfig:regression_usage}).
 
\paragraph*{S2 Text. Language links between items with many synonyms.}
\label{si:text:language link}

\begin{figure}[ht!]
    \centering
    \includegraphics[width=0.95\textwidth]{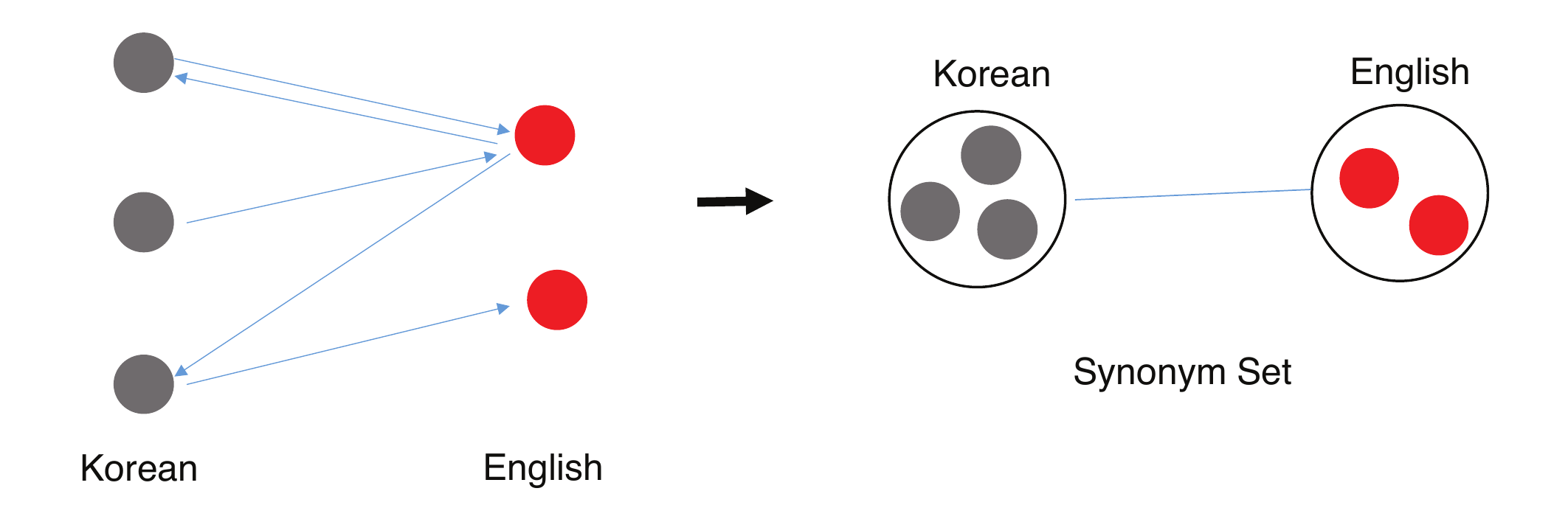}
    \caption{\textbf{Example of complex interlanguage links in Wikipedia}. Each node represents a scientific concept (\textit{i.e.} article/category), with the color of the node indicating the language edition each document belonged to, and the directed link denotes the existence of an interlanguage link between them.}
    \label{suppfig:synoym}
\end{figure}

Although most interlanguage links in Wikipedia are in a one-to-one relationship, there are few cases with many-to-many or one-to-many relationships in the dataset. As an illustrative example, we present a complex interlanguage relationship in Fig.~\ref{suppfig:synoym}. For such cases, we merged multiple items into a single node using the interlanguage links recursively until there were no more synonyms and removed the directionality of language links.

\paragraph*{S3 Text. Calculation of subject similarity for many-to-many case}
\label{si:text:complec_case}

\begin{figure}[ht!]
    \centering
    \includegraphics[width=0.65\textwidth]{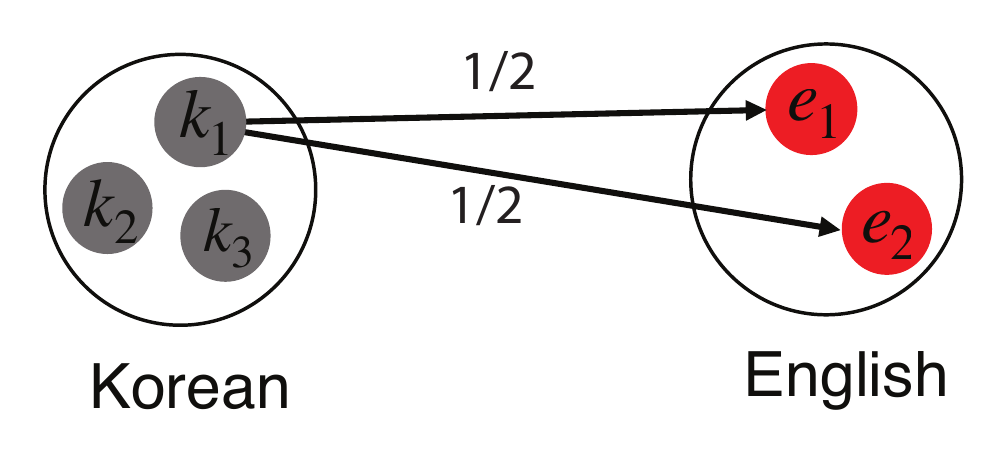}
    \caption{\textbf{Example of complex case of the interlanguage link records} Each node represents a scientific concept (\textit{i.e.} article/category), the color of the node indicates the language edition that each document belongs to,  and the directed link denotes the existence of an interlanguage link between them. For such many-to-many cases, we use a fractional sum to calculate the similarity between two languages.}
    \label{suppfig:mtomexample}
\end{figure}

In our main text, we only presented the simplest example to depict the computation of similarity when one subject is solely connected to another subject (one-to-one), whereas many-to-many cases, similar to the one displayed in Fig.\ref{suppfig:mtomexample}, exist. In this case, scientific concept $x$ in language edition $a$ is composed of sub-concepts $k_1, k_2,$ and $k_3$, and scientific concept $x$ in language edition $b$ is composed of sub-concepts $e_1$ and $e_2$. We first define the similarity of concepts using the fractional calculations as follows:

  \begin{equation}
      	d_{k_1,e_1}^{a \rightarrow b} = d(p_{k_1}^{a} T^{a \rightarrow b}, p_{k_2}^{b}),
  \end{equation}

\noindent where $p_{k_1}^{a}$ is a genealogy vector of subject $k_1$ of language edition $a$ and $p_{e_1}^{b}$ is a genealogy vector of subject $e_1$ of language edition $b$. Similiary, $T^{a \rightarrow b}$ is translation matrix between two different knowledge networks from language link data, while  $T^{a \rightarrow b}_{k_1e_1}$ is $\frac{1}{2}$ because $e_1$ and $e_2$ share the interlanguage relationship. Then, we convert distance into similarity  and define the subject similarity for many-to-many cases by aggregating the similarities among sub-concepts as follows:

  \begin{equation}
  s_{x}^{a \rightarrow b} = \frac{1}{N*M} \sum_{i=1}^N\sum_{i=1}^M s_{k_i,e_j}^{a \rightarrow b}
  \end{equation}

\noindent where $N$ and $M$ is the number of sub-concepts of $x$ in language edition $a$ and $b$, respectively. In the Fig.\ref{suppfig:mtomexample}, $N$ is 3 and $M$ is 2.

 \begin{figure}[ht!]
    \centering
    \includegraphics[width=0.95\textwidth]{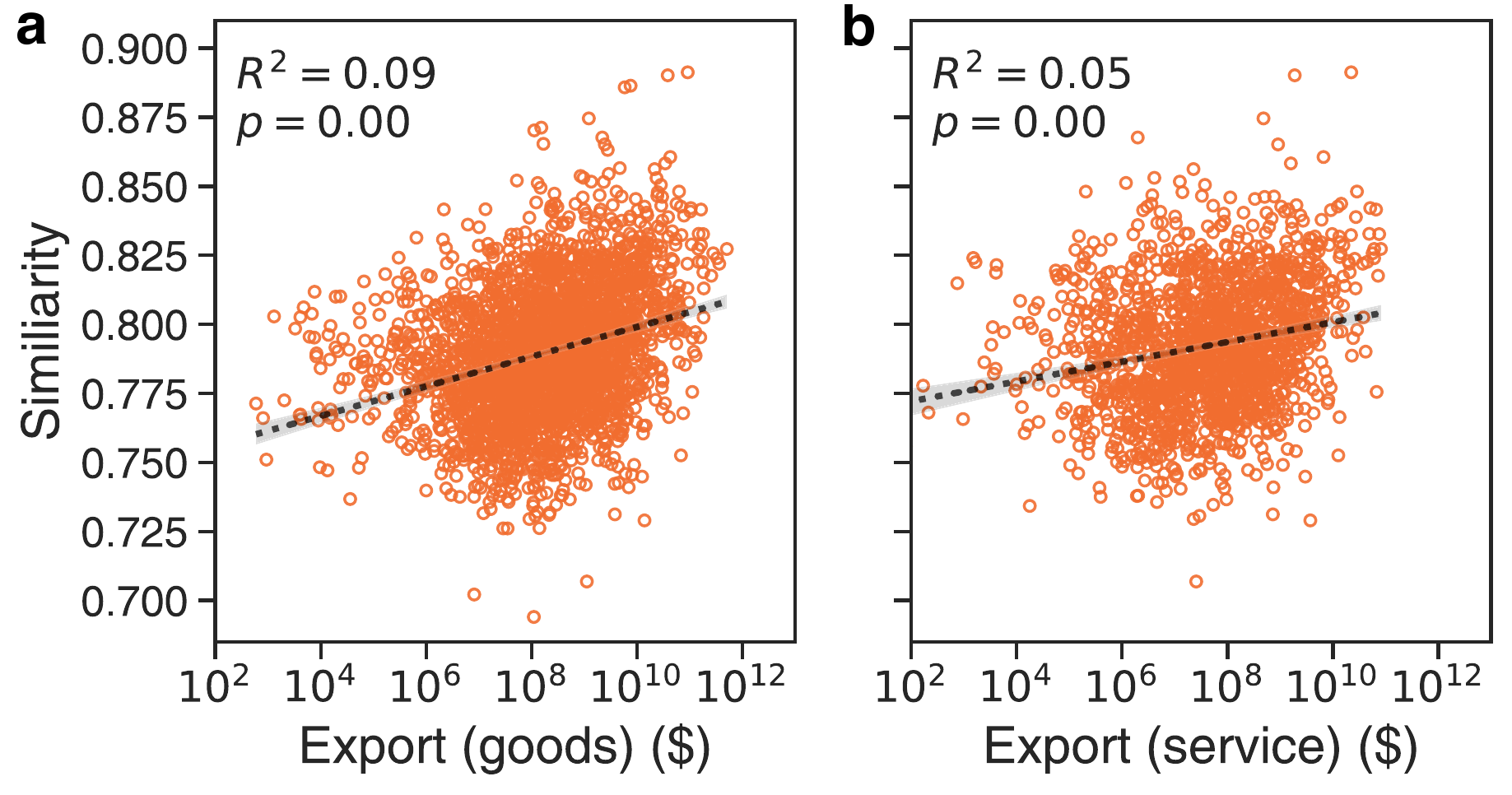}
    \caption{\textbf{Correlation between the structural similarity of the knowledge and export statistics by their types (goods, services) from UN Comtrade data.}  \textbf{a.} Physical goods. \textbf{b.} Services without physical goods.}
    \label{suppfig:good_service}
\end{figure}

 \begin{figure}[ht!]
    \centering
    \includegraphics[width=0.95\textwidth]{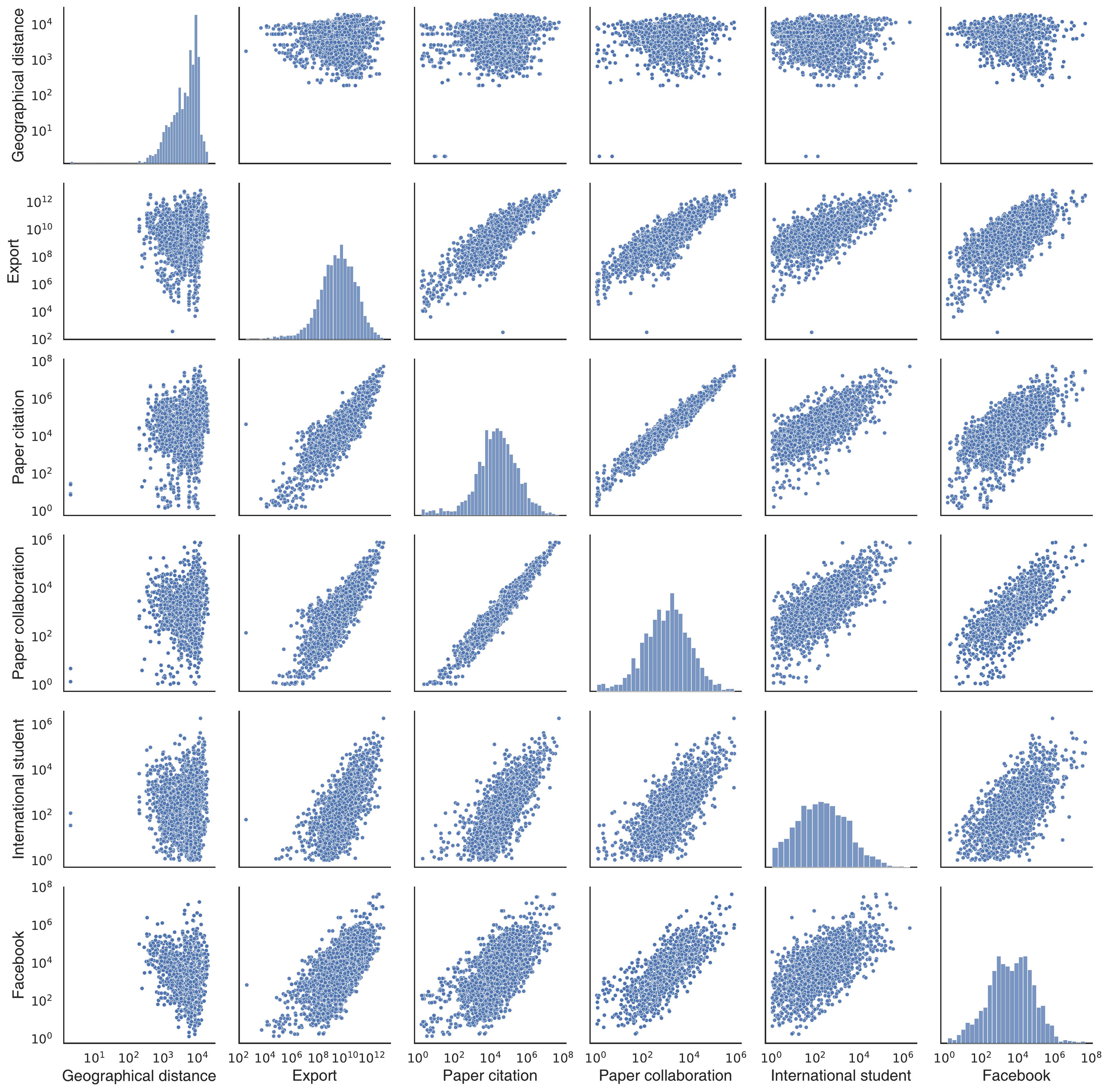}
    \caption{\textbf{Interrelations between socio-economic proximity} A diagonal figure shows a histogram of each proximity, and an off-diagonal figure indicates a scatter plot between proximities. We find that most socio-economic proximities are correlated, to some degree, except the geographical distance.}
    \label{suppfig:socio-economic parameters}
\end{figure}

\newpage

\begin{longtable}{llrr}

\caption{\textbf{Descriptive statistic of the constructed knowledge network by language}}
\label{supp:table:language_code}\\
\hline
\textbf{Code}    & \textbf{Name}  & \textbf{Number of Nodes} & \textbf{Number of links} \\
\hline
\textbf{af}      & Afrikaans      & 79,479                   & 242,563                  \\
\textbf{ar}      & Arabic         & 1,623,268                & 12,963,779               \\
\textbf{az}      & Azerbaijani    & 193,388                  & 858,047                  \\
\textbf{be}      & Belarussian    & 225,706                  & 947,555                  \\
\textbf{bg}      & Bulgarian      & 303,886                  & 1,333,211                \\
\textbf{bn}      & Bangla         & 115,403                  & 359,456                  \\
\textbf{bs}      & Bosnian        & 114,643                  & 351,321                  \\
\textbf{ca}      & Catalan        & 674,575                  & 2,089,419                \\
\textbf{cs}      & Czech          & 521,941                  & 1,942,054                \\
\textbf{cy}      & Welsh          & 124,742                  & 353,852                  \\
\textbf{da}      & Danish         & 291,853                  & 1,476,579                \\
\textbf{de}      & German         & 2,381,795                & 10,876,767               \\
\textbf{el}      & Greek          & 205,223                  & 999,238                  \\
\textbf{en}      & English        & 10,682,409               & 60,782,675               \\
\textbf{es}      & Spanish        & 1,709,975                & 6,338,496                \\
\textbf{et}      & Estonian       & 197,503                  & 471,143                  \\
\textbf{eu}      & Basque         & 354,017                  & 796,487                  \\
\textbf{fa}      & Persian        & 1,196,211                & 6,552,738                \\
\textbf{fi}      & Finnish        & 491,868                  & 1,381,502                \\
\textbf{fr}      & French         & 2,554,249                & 29,807,035               \\
\textbf{gl}      & Galician       & 206,363                  & 638,295                  \\
\textbf{he}      & Hebrew         & 299,628                  & 2,130,917                \\
\textbf{hi}      & Hindi          & 138,195                  & 265,514                  \\
\textbf{hu}      & Hungarian      & 475,530                  & 1,397,811                \\
\textbf{hr}      & Croatian       & 199,983                  & 450,381                  \\
\textbf{hy}      & Armenian       & 303,787                  & 1,446,520                \\
\textbf{id}      & Indonesian     & 597,187                  & 3,221,029                \\
\textbf{it}      & Italian        & 1,909,641                & 11,414,158               \\
\textbf{ja}      & Japanese       & 1,361,591                & 7,086,498                \\
\textbf{ka}      & Georgian       & 164,691                  & 405,990                  \\
\textbf{kk}      & Kazakh         & 226,944                  & 718,028                  \\
\textbf{ko}      & Korean         & 678,624                  & 3,480,057                \\
\textbf{lt}      & Lithuanian     & 210,819                  & 462,432                  \\
\textbf{lv}      & Latvian        & 112,990                  & 390,009                  \\
\textbf{ml}      & Malayalam      & 81,660                   & 249,272                  \\
\textbf{mr}      & Marathi        & 62,800                   & 140,629                  \\
\textbf{ms}      & Malay          & 333,024                  & 1,005,521                \\
\textbf{nl}      & Dutch          & 2,091,930                & 9,296,820                \\
\textbf{nn}      & Norwegian Nynorsk      & 201,295                  & 690,416                  \\
\textbf{no}      & Norwegian
     & 678,167                  & 4,020,650                \\
\textbf{pl}      & Polish          & 1,416,792                & 4,448,917                \\
\textbf{pt}      & Portuguese     & 1,311,643                & 6,537,942                \\
\textbf{ro}      & Romanian       & 507,569                  & 1,742,034                \\
\textbf{ru}      & Russian        & 1,800,296                & 10,236,325               \\
\textbf{sco}     & Scots          & 85,369                   & 393,524                  \\
\textbf{sh}      & Serbo-croatian & 460,908                  & 1,462,880                \\
\textbf{sk}      & Slovak         & 286,875                  & 1,004,445                \\
\textbf{sl}      & Slovenian      & 209,773                  & 793,713                  \\
\textbf{sr}      & Serbian        & 613,879                  & 1,608,806                \\
\textbf{sv}      & Swedish        & 4,879,164                & 19,527,411               \\
\textbf{ta}      & Tamil          & 144,399                  & 422,522                  \\
\textbf{th}      & Thai           & 155,711                  & 501,649                  \\
\textbf{tl}      & Tagalog        & 86,841                   & 334,451                  \\
\textbf{tr}      & Turkish        & 515,182                  & 2,288,882                \\
\textbf{uk}      & Ukrainian      & 1,057,032                & 5,611,258                \\
\textbf{ur}      & Urdu           & 464,973                  & 1,616,291                \\
\textbf{vi}      & Vietnamese     & 1,416,588                & 4,266,653                \\
\textbf{zh}      & Chinese        & 1,483,081                & 5,370,607                \\
\textbf{zh\_yue} & Cantonese      & 77,024                   & 222,048                  \\ \hline
\end{longtable}

\end{document}